\documentclass[fleqn,11pt]{wlscirep}

\usepackage{latexsym,amsmath}
\usepackage{hyperref}
\usepackage{float}
\usepackage{microtype}
\usepackage{bm}
\usepackage{amsbsy}
\usepackage{url}
\usepackage{graphicx}

\newcommand{\new}[1]{{\color{black}#1}}

\newcommand{\vect}[1]{\mbox{$\boldsymbol #1$}}

\newcounter{figcountSI}

\newcounter{tabcountSI}

\newcounter{seccountSI}

\begin{document}

\title{The structured backbone of temporal social ties}

\author[1]{Teruyoshi Kobayashi}
\affil[1]{Department of Economics, Center for Computational Social Science, Kobe University, Kobe, Japan}
\author[2]{Taro Takaguchi}
\affil[2]{Toda, Saitama 335-0021, Japan}


\author[3,4,*]{Alain Barrat}

\affil[3]{Aix Marseille Univ, Universit\'e de Toulon, CNRS, CPT,
  Marseille, France}
 \affil[4]{Data Science Laboratory, ISI Foundation, Torino, Italy}
\affil[*]{$\rm{Corresponding~author:~Alain.Barrat@cpt.univ}$-$\rm{mrs.fr}$}


\begin{abstract}
In many data sets, crucial information on the structure and temporality of a system coexists with noise and non-essential elements. 
In networked systems for instance, some edges might be non-essential or exist only by chance. Filtering them out and
extracting a set of relevant connections, the "network backbone", is a non-trivial task, and methods put forward until now do not address time-resolved networks, whose availability has strongly increased in recent years.
We develop here such a method, by defining an adequate temporal network
null model, which calculates the random chance of nodes 
to be connected at any time after controlling for their activity. This allows
us to identify, at any level of statistical significance, 
pairs of nodes that have more interactions than expected
given their activities: These form a backbone of
significant ties. We apply our method to empirical temporal networks
of socio-economic interest and find that (i) at given level 
of statistical significance, our method identifies more significant ties than 
methods considering temporally aggregated networks, and  
(ii) when a community structure is present, most significant ties 
are intra-community edges, suggesting that 
\new{the weights of}
inter-community edges 
\new{can be explained by the null model of random interactions}. 
Most importantly, our filtering method can assign 
a significance to more complex structures such
as triads of simultaneous interactions, while methods based on static
representations are by construction unable to do so. Strikingly, we uncover
that significant triads are not equivalent to triangles
composed by three significant edges. Our results hint at new ways
to represent temporal networks for use in data-driven models
and in anonymity-preserving ways.
\end{abstract}

\maketitle

\section{Introduction}

The analysis of large-scale empirical data sets, and in particular of complex networked data, is often made difficult by the nature of the data itself: 
data may be noisy \cite{butts2003error,newman2018noisy,newman2018noisyarxiv},
and contain both robust, generalizable properties and details 
specific to the collected data set under investigation, 
which would change if the data had been collected at a different moment or for a different sample of the same system. For instance, data describing
interactions between individuals (face-to-face~\cite{Cattuto2010PLOS,Stehle2011PLOS}, 
phone calls~\cite{JoKarsai2012NewJPhys,Schlapfer2014}, or online interactions~\cite{Centola2010Science,sapienza2018NTF}) might show robust group structures in different days or weeks
but with a different set and timing of interactions each day: 
the exact timing of an interaction in a specific day might not be relevant
to the understanding of the population's characteristics. Another issue might arise if the 
network under scrutiny is dense but with very heterogeneous weights on the edges. The importance of edges might then not be easily
reducible only to their own weight, nor to the local properties of the nodes they link, such as their degree (number of neighbours in the network)
or their strength (sum of the weights of their edges).

In order to extract the most relevant information from the data, several approaches
have been put forward for static networks. For instance, the k-core decomposition focuses on more and more connected parts of a network and
has been established as an important tool to analyze and visualize complex networks and to determine
influential spreaders in networks \cite{Seidman:1983,Alvarez:2005,Kitsak:2010}.
Another approach consists in determining a ``backbone'' of significant edges in the network, and to filter out the remaining non-essential edges.
Several methods have been proposed to this purpose in the case of static weighted networks. 
The simplest way of filtering edges is through thresholding: all the edges with weight below a given threshold value are removed.
Such a method however imposes an arbitrary cutoff scale,
while many systems of interest display broad distributions of weights and complex patterns at multiple scales. Other methods have thus 
been put forward to filter out edges simultaneously at different scales, using statistical tests based on null models~\cite{Serrano2009PNAS,tumminello2011statistically,li2014statistically,Hatzopoulos2015,gemmetto2017arxiv,casiraghi2017relational,marcaccioli2018parametric}: 
the fundamental idea is to test  whether the weight of an edge is distinguishable from the hypothetical one that would be generated at random
by a certain null model. Filtering is performed by fixing a desired significance level and selecting only those edges whose weight cannot be
explained by the null model at the chosen significance level. These significant edges form a backbone of the network.

Various null models have been proposed in the literature
to deal with static weighted networks \cite{Serrano2009PNAS,tumminello2011statistically,li2014statistically,Hatzopoulos2015,gemmetto2017arxiv,casiraghi2017relational,marcaccioli2018parametric}.  
The recent surge in the availability of temporally-resolved 
 high-resolution data on social and economic networks highlights however 
 the need for methods specifically designed to extract backbones from 
 temporal networks or temporally aggregated networks~\cite{Holme2012PhysRep,Masuda2016book}.
Obviously, each method defined on static weighted methods can be applied to a temporally aggregated network: for instance, a simple threshold
could be applied on the numbers of 
\new{events} 
\new{(i.e., on the weights of the aggregated edge)} 
between two nodes~\cite{grabowicz2014fast}. 
However, a highly active node could in principle have 
a large number of (non-essential) ties, so that one needs to control for 
 the difference in intrinsic activity levels across nodes to extract statistically significant ties that cannot be explained by random chance. 
 
 Here, we develop a method to extract an irreducible backbone from a sequence of temporal contacts between nodes, by 
 defining an adequate temporal null model. This null model can be interpreted as a (temporal) configuration or fitness model,
 whose parameters are estimated by using global information, namely the numbers of contacts for all node pairs,
 similarly to the enhanced configuration model (ECM) filter defined for static networks~\cite{gemmetto2017arxiv}. Thanks to this null model, 
we determine the set of significant ties, at any significance level, 
among all the pairs
of nodes \new{having interacted}. 
These ties form an irreducible backbone in the sense
defined in \cite{gemmetto2017arxiv}, as their significance cannot be reduced to the activity of the involved nodes. Most importantly, 
the temporal nature of the null model allows us to attribute a significance also to higher order structures such as e.g.,
simultaneously occurring triplets of interactions or other temporal motifs \cite{Kovanen2011JStatMech}, a task that would be by construction
impossible when defining significant ties and backbones directly from a temporally aggregated network. 

We \new{validate our filtering method on a synthetic benchmark and}
illustrate 
\new{its} application 
on temporal networks of social and economic relevance, \new{for which we} 
compare its results with several static filtering methods \new{and with a baseline
temporal extension of a static filter, obtained by simply applying
the same static filter to the successive snapshots 
of the temporal network}. 
Interestingly, at a given level of significance, our method 
\new{generally} identifies more significant
edges than other filters 
\new{and, as other filters based on null models, is able to detect
significant edges at all scales on interaction intensity}. 
Moreover, in cases where the aggregated network has a clearcut community structure,
corresponding e.g. to classes in a school, the significant ties turn out to be mostly intra-community ones: at high significance levels,
the network of significant ties (i.e., backbone) breaks into several connected components, each corresponding to one community, and inter-community edges turn out to be non-significant. 
This suggests that inter-community edges, while playing a crucial role in reducing the diameter of the network~\cite{Granovetter1973AJS,Watts1998Nature},
are here indistinguishable from randomly created edges, 
once nodes' activities are fixed. We also
\new{illustrate the ability of our filtering method to assign
a significance to higher order structures by investigating}
significant
triads, defined as sets of three nodes that interact simultaneously with each other more than expected given their activities.
\new{Strikingly, it turns out} that these
significant triads are not necessarily composed of three significant edges. This shows the crucial importance of 
taking into account temporality
when defining a null model to detect the significance of structures in temporal networks, as such information could not be obtained from 
a purely static null model \new{nor from the simple extension obtained
by applying a static filter to each temporal snapshot}.

\section{Data}

We consider \new{eight} data sets \new{describing systems of very different nature and} of social and economic interest, described by temporal networks (Table.~\ref{tab:data}). 
\new{The first} four data sets correspond to face-to-face contacts among individuals in different contexts, recorded using wearable sensors
by the SocioPatterns collaboration with a temporal resolution of 20 seconds
and publicly available~\cite{SocioPatterns}. We consider data sets collected in contexts with very different
activity levels, constraints on the schedule of individuals, duration and group structures, namely
a high school ("Highschool") \cite{Fournet:PLOS2015}, 
a primary school ("Primaryschool") \cite{Stehle2011PLOS}, 
an office building ("Workplace") \cite{Genois:2017} and 
a hospital ward ("Hospital") \cite{Vanhems:2013}.
\new{The fifth data set, ``Interbank'', is a temporal financial network} 
in which nodes and edges represent banks and overnight lending-borrowing relationships, respectively. 
Since overnight loan contracts last only for one day, we can construct a sequence of daily snapshot networks 
(i.e., time resolution is one day)~\cite{kobayashi2018social,kobayashi2018extracting}. 
We consider here the data on the online interbank market in Italy, called e-MID, 
between June 12, 2007 and July 9, 2007 (i.e., 20 business days). 
The data is commercially available from e-MID SIM S.p.A. based in Milan, Italy~\cite{emidHP}. 
\new{The sixth data set is the temporal network of emails exchanged between members of a European research institution (``Email")~\cite{SNAP}.
In the Email data, we consider daily network snapshots in which 
nodes represent individuals and an edge between two nodes
denotes the presence of at least
one email exchanged between the corresponding persons on a given day.
The seventh one describes the trips taken by customers of London Bicycle Sharing Scheme (``LondonBike")~\cite{LondonBike}. The nodes represent bike sharing stations and edges denote the presence of trips between two stations on June 22, 2014.  
Finally, the eighth data set is given by the UK domestic airline network
between 1990 and 2003 (``UK-airline")~\cite{UKairline}, at yearly
resolution, in which nodes 
and edges denote the UK airports and the presence of direct flights connecting 
two airports. 
For all datasets, we consider undirected networks. 
More details about data processing are provided in section ``Methods''.        
}

\begin{table}[tb]
    \caption{Basic description of empirical temporal networks. 
    {\rm 
    For the Interbank data, the number shown in the third column 
    denotes the number of daily edges rather than the total number of transactions.
    The ``\# communities" column gives the number of classes for the Primaryschool and Highschool data, of
    office departments for Workplace, and of types of occupations for the Hospital data. 
     We classify banks into Italian banks and foreign banks. \new{For the Email, LondonBike and UK-airline datasets, communities are detected by applying Infomap~\cite{Rosvall2008PNAS} on the aggregate weighted network. $Q$ is the weighted modularity of the corresponding partition.
     }}}
    \centering
    \begin{tabular}{lcccccc}
    \hline
    \hline
    Data            & $N$ & \# temporal edges & \new{\# aggregate edges} & Time span  & \# communities & \new{$Q$}\\ 
    \hline
    Highschool &  327  &  188,508  &5,818  & 5 days  &9 & 0.809\\
    Primaryschool  &   242  & 125,773  & 8,317   & 2 days  &10 & 0.627\\
    Workplace   &       217 &   78,249 & 4,274  & 10 days&  12 & 0.624\\
    Hospital     &     75  &   32,424   & 1,139 & 5 days &  4 & 0.215\\
    Interbank    &     162  & 7,104   &2,140  & 20 days & 2 & 0.036\\
    \new{Email} &      986     &   332,334 & 16,064 & 526 days  & $42$ & 0.669\\
    \new{LondonBike}     &  743  & 38,023 &  18,752 & 24 hours &  $5$ & 0.424\\
    \new{UK-airline}     &  55  & 2,787 & 398 & 14 years &  $3$ & 0.081\\
    \hline
    \end{tabular}
    \label{tab:data}
\end{table}

\section{Results}

\subsection{Temporal fitness model}

We consider a set of $N$ nodes and 
a sequence of interactions that occur at arbitrary points in time between these nodes~\cite{HolmeSaramaki2013book_Springer,Masuda2016book}. 
We fix a temporal resolution $\Delta$ by dividing the whole data temporal window of length $T$ into $T/\Delta$ time intervals, and we build
on each interval a binary adjacency matrix $A_t$ with elements $A_{ij,t}$ equal to $1$ if there is at least one interaction 
between $i$ and $j$ during  $(t -\Delta , t]$ and zero otherwise. 

In the temporal fitness model, each node $i$ is assigned an \new{intrinsic variable that we call} ``activity'' level, 
$a_{i}  \in  (0,1]$, and the probability $u$ that nodes $i$ and $j$ interact 
 (e.g., through a face-to-face contact, 
 a bilateral financial transaction, etc) during any 
 given time interval is simply given by the product of their activity levels~\cite{kobayashi2017significant}:
\begin{align}
u(a_{i},a_{j}) = a_{i} a_{j},    \label{eq:matching_func}
\end{align} 
In a static network context, this class of network model is called the \textit{fitness model} and has been used to model network generative processes~\cite{Caldarelli2002PhysRevLett,Boguna2003PhysRevE,DeMasi2006PRE}. 
The null model we obtain is thus a sequence of successive independent realizations of the (static) fitness model. It settles a baseline of how much two nodes are expected to interact, given their activities, if interaction partners are selected at random at each time step. 
\new{In other words, we do not assume any underlying pre-existing network 
structure and
we consider a hypothetical situation in which there is always a positive chance of interaction between any two nodes.}
Note also that this 
temporal null model does not contain any a priori knowledge
of the group structure of the nodes. An interesting 
\new{modification} could
be to superimpose group labels or node properties (e.g., gender or age
for nodes representing individuals) and interaction probabilities
depending on the nodes' properties.

In the simplest version of this null model, we consider constant activity values for each node.
In addition, we present in section~\ref{sec:SI_timevarying} a refined version that takes into account temporal variations of the overall 
interaction activity in the system, through the introduction of a time-varying parameter $\xi(t)$. In that case, the null model is defined by the fact that the probability of nodes $i$ and $j$ establishing a connection at time $t$ is $u(a_{i},a_{j},t) = a_{i} a_{j} \xi(t)$. 
We present
here the case of constant $\xi(t)=1$ as we can then obtain an analytical form for the probability distribution of the number of interactions of a 
pair of nodes, while only an approximate formula is available in the more general case. 
Note that this number is at most $\tau = T/\Delta$, i.e., 
the number of time intervals given the resolution $\Delta$.
We show in section~\ref{sec:SI_timevarying} that both methods yield \new{quite similar} 
results in the case studied here.

\subsection{Significant ties}
\label{sec:sigties}

\begin{figure}[t]
\begin{center}
\includegraphics[width=.7\columnwidth]{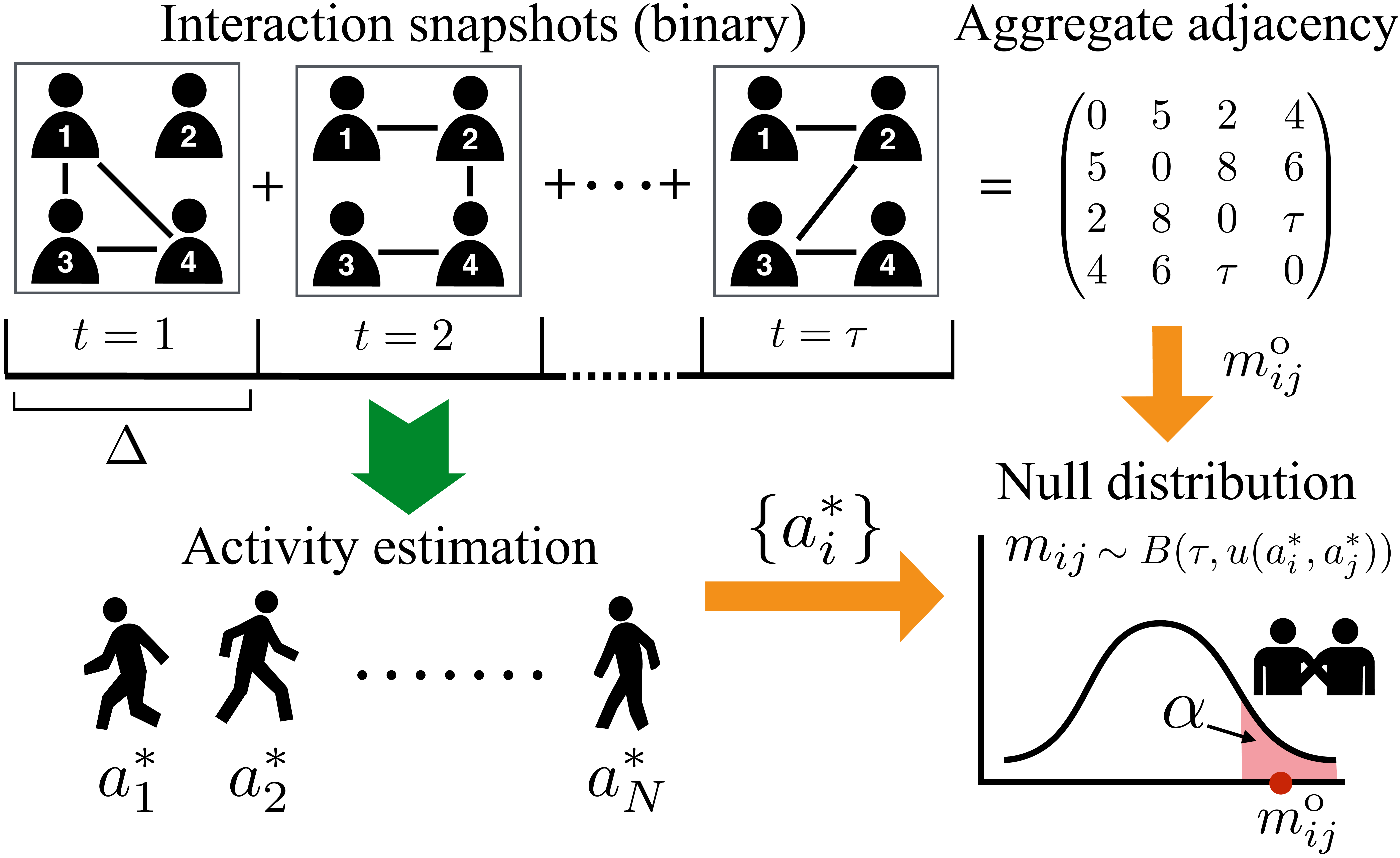}
    \end{center}
    \caption{Sketch of the filtering method. From the temporal network at resolution $\Delta$, described by
    $\tau=T/\Delta$ adjacency matrices, we estimate the set of node activities $(a_1^*,\ldots , a_N^*)$, and 
    thus the probability distribution of the number of interactions between any pair of nodes $(i,j)$ under the null model. We compare the
    empirical value $m_{ij}^{\rm o}$ with the percentiles of this distribution to determine the significance of the pair $(i,j)$'s interactions.}
 \label{fig:schematic}
\end{figure}

To uncover significant ties with respect to the null model described above, we proceed in two steps (Fig.~\ref{fig:schematic}).
\new{First, given a data set, we estimate the node
activity levels $\vect{a}\equiv (a_1,\ldots , a_N)$,
within the temporal fitness model. 
Note that the activities $\{a_i\}$ are latent variables that rule the probabilities of interactions between nodes in the model
but are not directly observable 
nor inferred from the local information about 
the nodes in a data set. They can however be estimated
using a maximum likelihood estimation}
as described in section~\ref{sec:activity_estimation}, yielding the values $\vect{a}^*\equiv (a_1^*,\ldots , a_N^*)$.
\new{From the definition of the model, the $a_i^*$ are 
expected to be correlated with the total number of interactions with 
other nodes, and we show indeed in 
the Supporting Information (SI)
that these estimated activity parameters turn out to be proportional 
to the strengths of the nodes and correlated with their degree.}

We then compute for each \new{interacting} pair of nodes $i$ and $j$ the probability distribution of their total number of interactions $m_{ij}$
in the null model, which is given by the following binomial distribution: 
 \begin{align}
   & g(m_{ij}|a_{i}^*,a_{j}^*) \notag \\
   &= \begin{pmatrix} \tau \\ m_{ij}\end{pmatrix} u(a_{i}^*,a_{j}^*)^{m_{ij}} (1-u(a_i^*,a_j^*))^{\tau-m_{ij}} . 
   \label{eq:edge_test}
\end{align} 
Let $m_{ij}^{c}$ denote the $c$-th percentile $(0 \leq c \leq 100)$ of $g(m_{ij}|a_{i}^*,a_{j}^*)$,
i.e., $c/100 = G(m_{ij}^{c} |a_{i}^*,a_{j}^*)$, where $G$ is the cumulative distribution function (CDF) of $g(m_{ij}|a_{i}^*,a_{j}^*)$\new{, namely  $G(m_{ij}^c|a_i^*,a_j^*) = \sum_{m_{ij}=0}^{m_{ij}^c}g(m_{ij}|a_i^*,a_j^*)$}. 
If the actual empirical number of interactions $m_{ij}^{\rm o}$
between $i$ and $j$ is larger than $m_{ij}^{c}$, it means that this
\new{empirical} number
cannot be explained by the null model at significance level 
$\alpha \equiv 1-c/100$, indicating that $i$ and $j$ are 
connected by a \textit{significant tie}. \new{The $p$-value of the test is given by $1-\sum_{m_{ij}=0}^{m_{ij}^{\rm o}}g(m_{ij}|a_i^*,a_j^*)$.}
For a given significance level $\alpha$, we can test the significance of 
the set of interactions 
\new{composing}
each 
\new{tie} independently from the other ties \cite{gemmetto2017arxiv}. Note that, even if the  significance of a tie is determined 
from an aggregated number of interactions, 
a significant tie does not correspond here to a static
edge but to an interacting pair of nodes with their
set of temporally resolved interactions, and
the backbone given by all interactions in the significant ties remains a temporal network.
Tuning $\alpha$ allows us to probe more and more significant pairs by decreasing $\alpha$, and/or to tune the number of ties retained in the backbone,
providing a systematic filtering method that we call Significant Tie (ST) filter.

Thanks to the use of the null model, a pair of interacting
nodes can be significant even if their number of interactions is small, 
as long as their individual activity levels are sufficiently low. Reciprocally, ties with a large number of interactions might
not be significant if the two involved nodes are very active. 
The ST filter controls indeed for the 
difference between nodes in terms of intrinsic activity levels. As a consequence, 
the significant ties identified by our method are ``irreducible" in the sense that their significance cannot be attributed to local node-specific 
properties~\cite{gemmetto2017arxiv}, such as the node degree and strength in the aggregated network: 
the probability of interaction between two nodes under the null hypothesis is determined by an interplay of global
and local information through the maximum likelihood estimation (Eq.~\eqref{eq:ML_a} in section~\ref{sec:activity_estimation}).
The resulting network of interactions between the significant pairs of nodes 
may thus be regarded as an \textit{irreducible backbone} of the temporal network under study~\cite{gemmetto2017arxiv}. \new{The Matlab code for the ST filter is available from a Zenodo website\cite{STcode}.}

 \subsection{\new{Beyond significant ties: significant temporal structures}}

Using a temporal fitness model as null model allows us to go beyond the usual tests concerning the significance of ties,  
and to assign
a significance to higher order structures such as temporal motifs. To illustrate this point, let us consider the simple case of a triadic interaction
between three nodes $i$, $j$, $k$; the empirical number of time intervals in which the three pairs $(i,j)$, $(j,k)$ and $(i,k)$
are \textit{simultaneously} interacting, denoted by $r_{ijk}^{\rm o}$, can be compared to the probability distribution of the number $r_{ijk}$ of occurrences of such 
triangles in the null model.
For each time interval, the probability that $i$, $j$, $k$ are forming a triangle of interactions in the temporal fitness model is
\begin{align}
    v(i,j,k) = u(a_i^*,a_j^*)\cdot u(a_j^*,a_k^*)\cdot u(a_k^*,a_i^*) \, ,
\end{align}
so that $r_{ijk}$ obeys the following probability distribution in the null model:
\begin{align}
   &h(r_{ijk}|a_{i}^*,a_{j}^*,a_{k}^*)  \notag \\
   & = \begin{pmatrix} \tau \\ r_{ijk}\end{pmatrix} v(i,j,k)^{r_{ijk}} (1-v(i,j,k))^{\tau-r_{ijk}} .
\label{eq:triangle_test}
\end{align} 
Similarly to the case of dyads, we define for each significance level $\alpha=1-c/100$ the significant triads as those such that $r_{ijk}^{\rm o}$ 
is larger than the $c$-th percentile of $h(r_{ijk}|a_{i}^*,a_{j}^*,a_{k}^*)$.

Note that this method can easily be generalized to any set of temporally constrained interactions (e.g., occurring in a sequence of successive
snapshots) or motifs. On the contrary,
any filtering method based directly on the aggregated network, and not taking into account the temporality of the data, is by construction
unable to define a null hypothesis for simultaneous interactions (or interactions with a given temporal sequence) and thus to assign 
a significance to such patterns.

\subsection{\new{Validation and case studies}}

\new{We now apply the ST filter to both synthetic and real data sets,
and compare its outcome with other filtering methods. On the one hand,
we consider two methods that use directly the static, temporally aggregated network, namely the disparity filter (DP filter) \cite{Serrano2009PNAS} and the enhanced configuration model (ECM filter) \cite{gemmetto2017arxiv}, whose computation is recalled in section~\ref{sec:SI_DP_ECM} of the SI.
In addition, we also examine a method that partially 
takes into account the temporal nature of interactions by
implementing a static filter on each temporal snapshot.
Specifically, we apply the ECM filter on each snapshot, 
and a pair of nodes is regarded as significant 
if an edge between the two nodes is identified as 
significant in at least one snapshot: this defines a baseline
temporal filter that we call ECM-R (ECM-repeated).}

\subsubsection{\new{Synthetic networks}}\label{sec:synthetic}
\new{
Let us first consider as a validation exercise a synthetic
temporal network with known properties, composed of a superposition
of random and strong edges. We consider $N=300$ nodes, each node $i$
endowed with an internal variable 
$a_i^{\prime}\in [0,1]$ drawn from a Beta distribution, and
nodes $i$ and $j$ are connected at each time step 
with probability $a_i^{\prime}a_j^{\prime}$. We then 
superimpose to this random temporal structure additional 
temporal edges during a time-window of length $T$: 
we first select at random 20\% of the interacting pairs,
and we add interactions for these pairs, so that 
they become a known set of ``strong'' ties. 
We then treat these synthetic data as the other data sets: we
create a sequence of $\tau=T/\Delta$ network snapshots by 
aggregating all the interactions made over time-windows
of length $\Delta$. 
More details about the generation of synthetic networks are 
provided in section~\ref{sec:synthetic_process} of the SI.

  \begin{figure}
     \centering
     \includegraphics[width=.98\columnwidth]{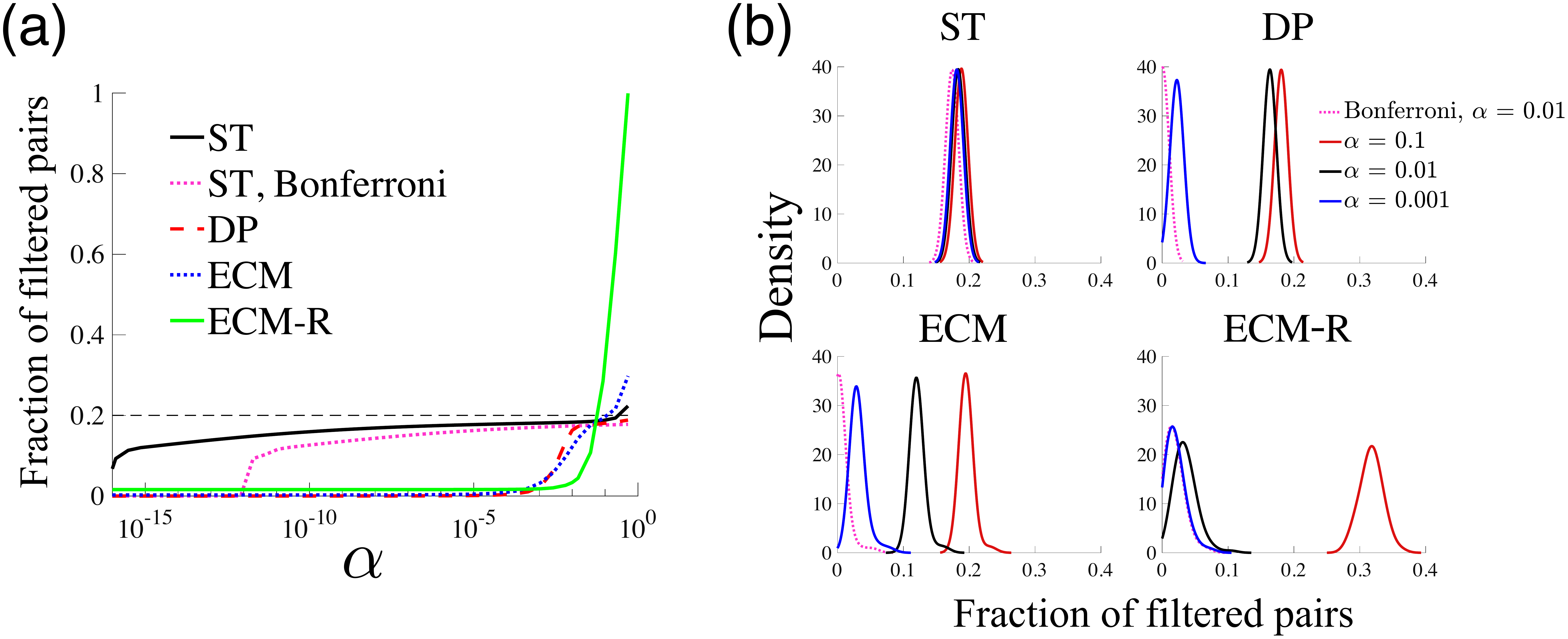}
     \caption{\new{Fraction of significant pairs detected in synthetic temporal networks. Synthetic networks are generated such that exactly $20\%$ of edges have strong ties. We generate 100 networks 
     with $N=300$, $T=300$ and we use $\Delta=10$ (See section \ref{sec:synthetic_process} of the SI for a detailed description of synthetic networks). (a) Average fraction of significant ties as a function of the significance level. The dotted pink line denotes ST edges with Bonferroni correction for multiple tests.  
     (b) Distribution of the fraction of significant pairs 
     for different significance levels.  
     }}
     \label{fig:synthetic_net}
 \end{figure}

Figure~\ref{fig:synthetic_net}{\sf a} shows that the
fraction of significant edges detected by all the filters we consider
decreases as $\alpha$ decreases. 
and lies below the fraction of known strong pairs (namely $0.2$) 
as soon as $\alpha$ takes reasonable values such as $\alpha<0.01$.
The fraction of false-positives is then
negligible in all cases (see Fig.~\ref{fig:synthetic_excl} in the SI), 
almost all ties detected as significant correspond indeed to strong 
edges in the synthetic data (for $\alpha<0.01$). Moreover, 
the ST filter is much more successful in detecting 
strong ties compared to the other filtering methods examined as
soon as $\alpha$ decreases below $0.01$:   
Figure~\ref{fig:synthetic_net} clearly shows that both the
average and the distribution of the fraction of strong edges
detected by the ST filter are stable for a broad range of $\alpha$, 
while the fraction of detected strong ties decreases very fast 
for the other filters when the significance level increases.
Note that no filter detects all strong ties. This is due
to the fact that some of the node pairs selected to be ``strong" ties
connect nodes with large activity values: the additional
interactions might then yield an overall temporal sequence
that is still compatible with the null model, i.e., their number of
interactions could still be explained by chance, given their activity
levels. This means overall that it is reasonable to regard the 
fraction of 
ST edges as a conservative measurement for the fraction of significant
ties in a data set. 
}

\new{Let us now turn to the empirical data case studies.  
We present in the main text the main results for a subset of the
data sets considered and refer to the SI for the other data sets.}

 \subsubsection{Number of significant ties}

\begin{figure}[t]
\begin{center}
               \includegraphics[width=.9\columnwidth]{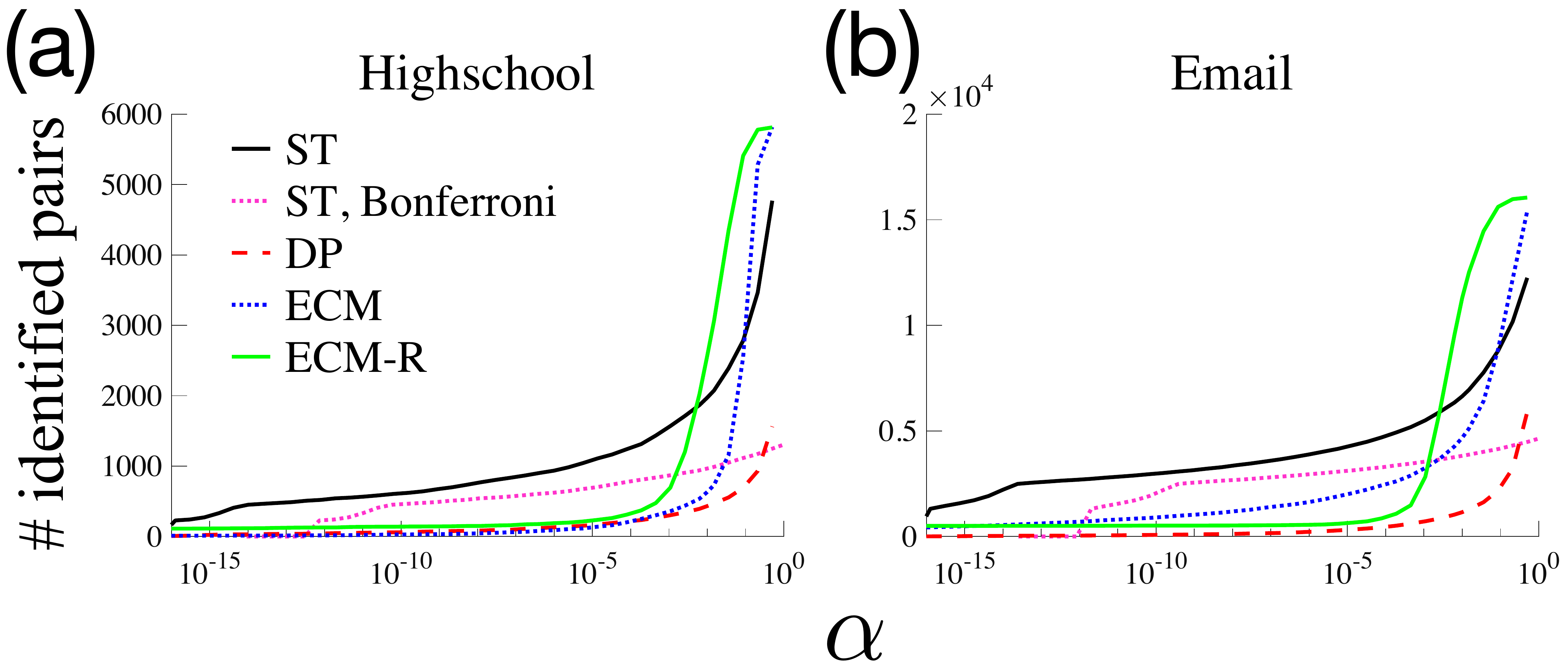}
    \end{center}
    \caption{\new{Number of significant ties as a function of the significance level for the (a) Highschool and
    (b) Email data sets. The dotted pink line denotes the ST filter with Bonferroni correction, while black solid, red dashed, blue dotted and green solid lines represent the ST, DP, ECM and ECM-R filters without Bonferroni correction, respectively. Temporal resolution is $\Delta = 15$ min for Highschool and 1 day for Email.
  }}
 \label{fig:number}
\end{figure}

Figure~\ref{fig:number} first displays the number of significant ties as a function of the significance level $\alpha$, for the 
\new{four} methods. As $\alpha$ decreases, this number decreases sharply for all methods. Interestingly however, 
the number of significant ties remains much larger for our method 
as soon as $\alpha$ enters a regime of high statistical significance (e.g.,  $\alpha < 10^{-2}$). 
As $\alpha$ becomes very small, i.e., at very high statistical significance, DP, ECM \new{and ECM-R filters} retain only a very small number of edges, while the ST filter still uncovers a relatively large number of significant node pairs. 
\new{Note that the ECM-R filter, which partially takes into account
temporality, tends to retain more significant ties than the static ECM
filter.
We also present the results for the ST filter with Bonferroni correction, in which the significance level is adjusted by dividing by 
the number of edges to control for type-I errors.} 
The resulting backbone of interactions between 
 significant pairs at very low $\alpha$ (e.g., $\alpha$ between $10^{-10}$ and $10^{-5}$) might be regarded as a 
 \textit{fundamental backbone} of the data set. See Fig.~\ref{fig:num_sigties_SI} for the similar
 results obtained with other data sets and different parameters.

\subsubsection{Comparison with other filtering methods} 

\begin{figure}[t]
\begin{center}
              \includegraphics[width=.9\columnwidth]{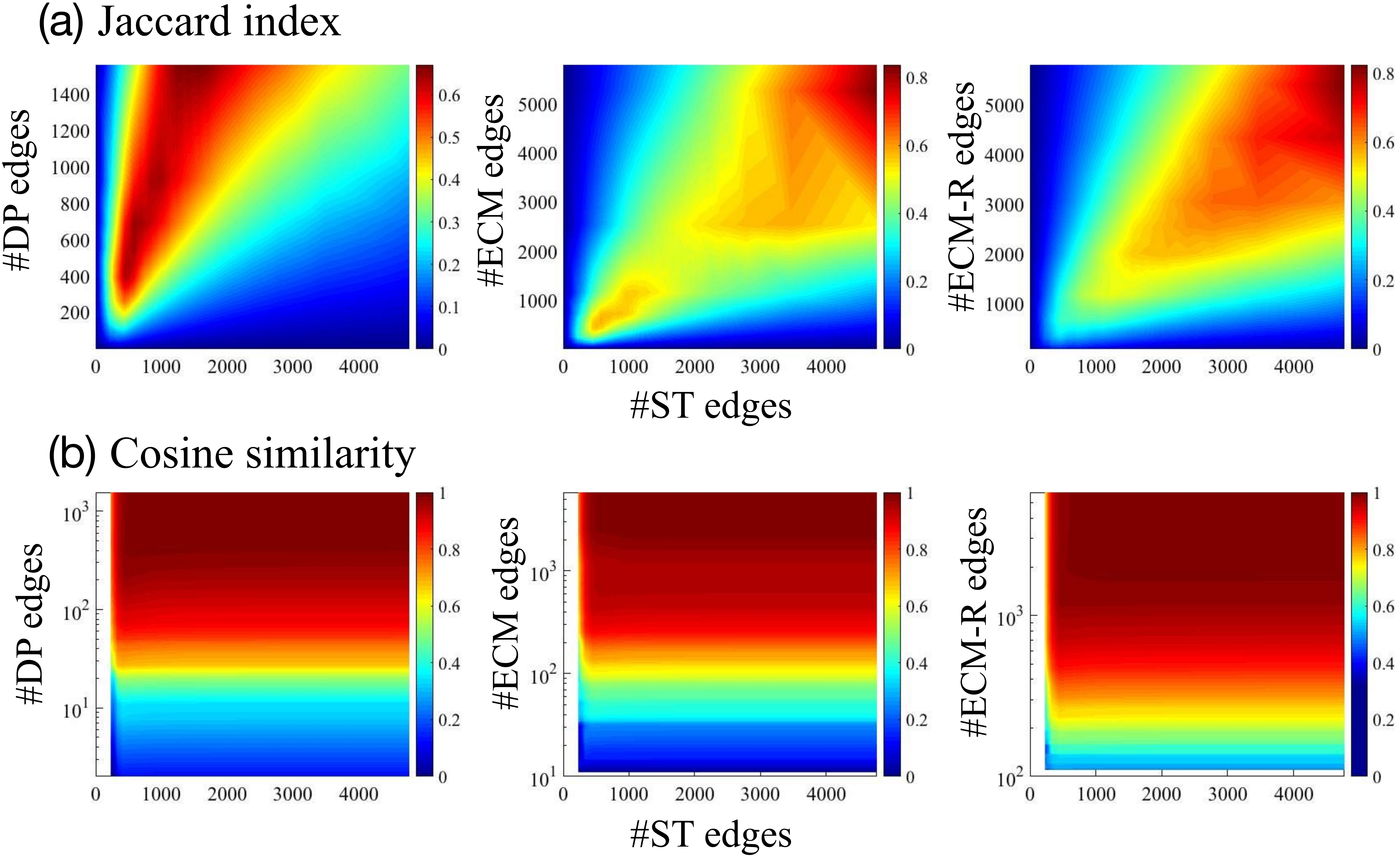}
             \end{center}
\caption{\new{Comparison between the sets of edges detected as 
significant by different filters, for the Highschool data. 
 (a) The Jaccard index is defined by $J(I_{\rm ST}^\alpha,I_{\rm x}^{\alpha'}) = {|I_{\rm ST}^\alpha \cap I_{\rm x}^{\alpha'}|}/{|I_{\rm ST}^\alpha \cup I_{\rm x}^{\alpha'}|}$, where $x\in\{{\rm DP},{\rm ECM},{\rm ECM{\text{-}}R}\}$. 
 We plot the Jaccard index versus the numbers of edges detected 
 at different significance levels, ranging from $\alpha = 10^{-17}$ to $0.5$ (obviously, the higher the value of $\alpha$, the larger the 
 number of significant edges). (b) Same for the cosine similarity 
 defined in Eq.~\ref{eq:cosine}.}}
 \label{fig:jaccard_highschool}
\end{figure}

Given the differences in the definition of the various filters, 
it is important
to understand to what extent these filters select distinct or similar sets of ties. 
\new{We first quantify the similarity 
between backbones obtained by different filters through} 
the Jaccard index 
$J(I_{\rm ST}^\alpha,I_{\rm x}^{\alpha'}) = \frac{|I_{\rm ST}^\alpha \cap I_{\rm x}^{\alpha'}|}{|I_{\rm ST}^\alpha \cup I_{\rm x}^{\alpha'}|}$ 
\new{that} gives the fraction of common edges
between the backbone obtained by the ST filter
at significance level $\alpha$ and another 
backbone ${\rm x}\in\{{\rm DP},{\rm ECM}, {\rm ECM\text{-}R}\}$ at significance level $\alpha'$. A Jaccard equal to $1$ means that both methods
yield the same exact set of edges, while $J=0$ means that the backbones are disjoints. 
As the different methods yield very different backbone sizes for a fixed significance level, we show in Fig.~\ref{fig:jaccard_highschool}{\sf a}
a color plot of the Jaccard index as a function of the number of node pairs retained by each filtering method. 
In all cases, the largest Jaccard indices are obtained when the number of edges are similar: they reach at most of $\sim 80\%$ \new{when $\alpha$ takes a meaningful value (e.g., $\alpha<10^{-2}$)} and decrease as the backbone size decreases (Fig.~\ref{fig:Jaccard_SI}), \new{showing
that the backbones obtained by different methods show some
similarity but are not equivalent}.

\new{To investigate this in more details, we also consider
a weighted measure of the similarity, as}
the Jaccard index 
does not take into account 
that different ties can correspond to very different number of interactions (i.e., weights). 
 \new{We thus compute the cosine similarity between the backbones 
 obtained by different filters:
\begin{align}
\sigma({\rm x},{\rm x'}) = \frac{\sum_{i<j}w_{ij}^{{\rm x}} w_{ij}^{{\rm x'}}}
  {\sqrt{\sum_{i<j}\left(w_{ij}^{{\rm x}}\right)^2}\sqrt{\sum_{i<j}\left(w_{ij}^{{\rm x'}}\right)^2}},
  \label{eq:cosine}
\end{align}
where ${\rm x}$ encodes both the filtering method (ST, DP, ECM, ECM-R) and the significance level $\alpha$ and the sums run
only on the pairs of nodes present in the backbones.
Figure~\ref{fig:jaccard_highschool}{\sf b} indicates that values larger than the Jaccard index are obtained, with similarities of $0.9 - 1$, decreasing below $0.8$ only when the backbone
sizes becomes small.} 
The various backbones seem thus to all retain similar sets of ties with large weights, while differing when assessing the significance of pairs of nodes with smaller numbers of interactions.
Similar results are obtained with all data sets 
(Fig.~\ref{fig:cosine_SI} in the SI).

\new{We explore further} in 
Fig.~\ref{fig:weight_dist} the distributions of weights (i.e., of the number of interactions) of the ties considered either as significant or not by different filters. Significant ties display on average larger weights than the non-significant ones. 
The DP filter in particular is almost equivalent to a thresholding procedure, 
\new{(i.e., selecting edges with high weights)}, 
with only a narrow range of weight values for which both significant and non-significant edges can be found. This is in agreement with the result
noted in \cite{gemmetto2017arxiv} that this filter tends to retain larger weights.
\new{For all the other filters, the weight distributions of the significant edges
are quite similar and, most importantly, are as broad as the original
weight distribution of the whole network: thanks to the use of null models, 
these filters manage to find significant ties at all intensity scales, i.e., 
with a wide range of numbers of interactions. For instance, for the ST filter
there exists a significant tie at $\alpha=0.01$ for the Workplace data set with
only a single interaction ($m_{ij}^{\rm o}=1$), and in other data sets 
some significant ties have $m_{ij}^{\rm o}=2$ or $3$ only.
}


 \new{We finally investigate the effect of varying the temporal resolution $\Delta$ in Fig.~\ref{fig:resol_heatmap} in the SI. 
 An increase in $\Delta$ (i.e., lower resolution) generally lowers the
 number of significant ties (Fig.~\ref{fig:resol_heatmap}{\sf a}), 
 reducing the Jaccard index between the sets of significant ties 
 obtained at different temporal resolutions as the difference in $\Delta$ increases (Fig.~\ref{fig:resol_heatmap}{\sf b}). 
 However, Fig.~\ref{fig:resol_heatmap}{\sf c}) shows that
 this decrease is mainly caused by the fact that some  
 significant pairs are no longer detected as $\Delta$ increases, 
 while ties detected as significant at a lower resolution 
 $\Delta'$ are also
 detected as significant at a higher resolution $\Delta$ ($\Delta < \Delta'$).
 }

\begin{figure*}[!h]
\begin{center}
              \includegraphics[width=.95\columnwidth]{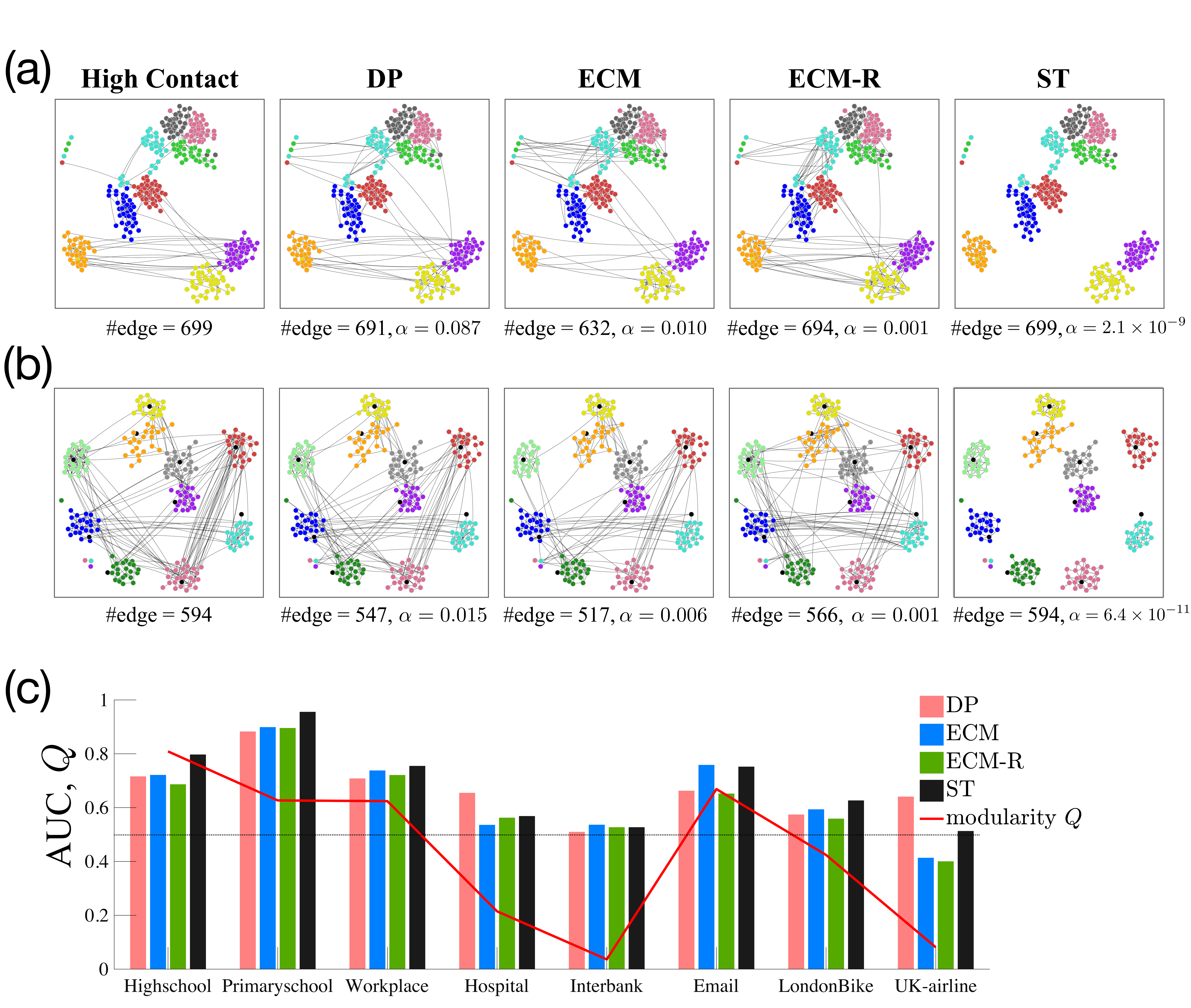}
    \end{center}
    \caption{Backbones and community structure. Visualization of the backbones obtained by different filtering methods, at similar numbers of edges, for the (a) Highschool and (b) Primaryschool data sets.    
    The nodes are shown in the same position for all the backbones of a given data set.
    "High contact" correspond to a simple thresholding procedure on the aggregated network.
    Different colors denote different classes. For the Primaryschool data, black circles represent the teachers. 
    The ST filter detects a larger fraction of intra-class edges than the other filtering methods.
    (c) AUC of the ROC curve for the identification of intra-community edges.  
    The red solid curve gives the weighted modularity $Q$~\cite{Newman_Girvan2004PRE,newman2004analysis} calculated by regarding the actual groups (classes for Primaryschool and Highschool, departments for Workplace,
    roles for Hospital) as communities in the original aggregate networks. 
    For the Interbank data set, banks are classified into Italian banks and foreign banks. \new{For Email, LondonBike and UK-airline, the communities are detected by Infomap~\cite{Rosvall2008PNAS}.}
    }
 \label{fig:fixed_edge_visual}
\end{figure*}

 \subsubsection{Backbones and community structure}
 
 In \new{several of}
 the data sets we consider, nodes can be classified into different groups, corresponding e.g. to classes in schools, departments 
 in the workplace and roles in the hospital. In the Highschool, Primaryschool and Workplace cases, these groups 
 define a clear-cut community structure 
 \cite{Stehle2011PLOS,Fournet:PLOS2015,Genois:2017}, while nodes from different 
 groups are more mixed in 
the 
\new{Hospital} data set \cite{Vanhems:2013}. 
This is confirmed by the values of the weighted modularity of the partition corresponding to these
groups shown in Fig.~\ref{fig:fixed_edge_visual}{\sf{c}} 
\new{and Table \ref{tab:data}}. 
\new{We also obtain a clear community structure for the Email and Londonbike cases.}

A visualization of the backbones obtained by the different filtering methods for the Highschool and Primaryschool data sets, 
shown in Fig.~\ref{fig:fixed_edge_visual}{\sf a} and {\sf b},
indicates that the backbone obtained by the ST filter seems to separate the network into connected components corresponding to these communities
more efficiently than the other filters, at fixed number of edges. 
We show that this is indeed the case through two quantitative indicators. 
First, we measure, as a function of the
backbone size, the fraction of intra-group edges (Fig.~\ref{fig:clust}). It is larger than the random baseline (in which edges are kept completely at random) for all
filters, approaching one as the number of edges decreases, and 
maximal for the ST filter. Second, we consider each filtering method as a prediction task for
finding intra-group edges: we use $\alpha$ as parameter and measure, for each $\alpha$, the true and false positives (edges in the backbone
that are/are not intra-group edges)  and true and false negatives (edges not in the backbone that are not/are intra-group edges), building thus
a ROC curve (see section~\ref{sec:SI_ROC} of the SI 
for details). The area under the curve (AUC)
of the ROC curve of the ST filter is higher than for the other filters for the 
\new{five} data sets for which 
\new{the modularity is high.}

The fact that the ST backbone selects mostly intra-community node pairs suggests that inter-community \new{interactions may be explained by the
null model of random interactions ruled by the node 
activity levels.
}
Inter-community edges, which act as bridges, play an important role in propagating information, spreading ideas, and diffusion of influence~\cite{Granovetter1973AJS,Onnela2007PNAS}. Nevertheless, our analysis shows that the actual \new{weights of these} 
edges are statistically indistinguishable from \new{the ones resulting from interactions at random times, 
regardless of the history of interactions between the two nodes,
once the node activities are given.}
This hints at a way to represent the original temporal network as a superposition of 
(i) a backbone of significant ties and (ii) connections
extracted at random \new{according to the null model} 
between nodes of different groups, in a way that would
refine the contact matrix of distributions put forward in \cite{Machens:2013,genois2015compensating}.

\new{While the ST filter detects intra-community edges more efficiently than the other filters do, our results show that 
the other filters also tend to detect
intra-community edges when there is an explicit community structure.
We investigate this further in the SI (Fig.~\ref{fig:share_sig_inter})
in which we show that the intra-community edges of a node are more likely 
to be significant compared to its inter-community edges, and this
for any node degree. This property is common to all filters but
most evident in the ST case. 
This suggests that the non-significance of inter-community edges 
is not explained by differences in aggregate degree of their 
end nodes, but rather reflects an intrinsic difference between 
intra- and inter-community edges. 
The ST filter is able to exploit such an intrinsic difference most efficiently.}

\begin{figure}[t]
\begin{center}
\includegraphics[width=.95\columnwidth]{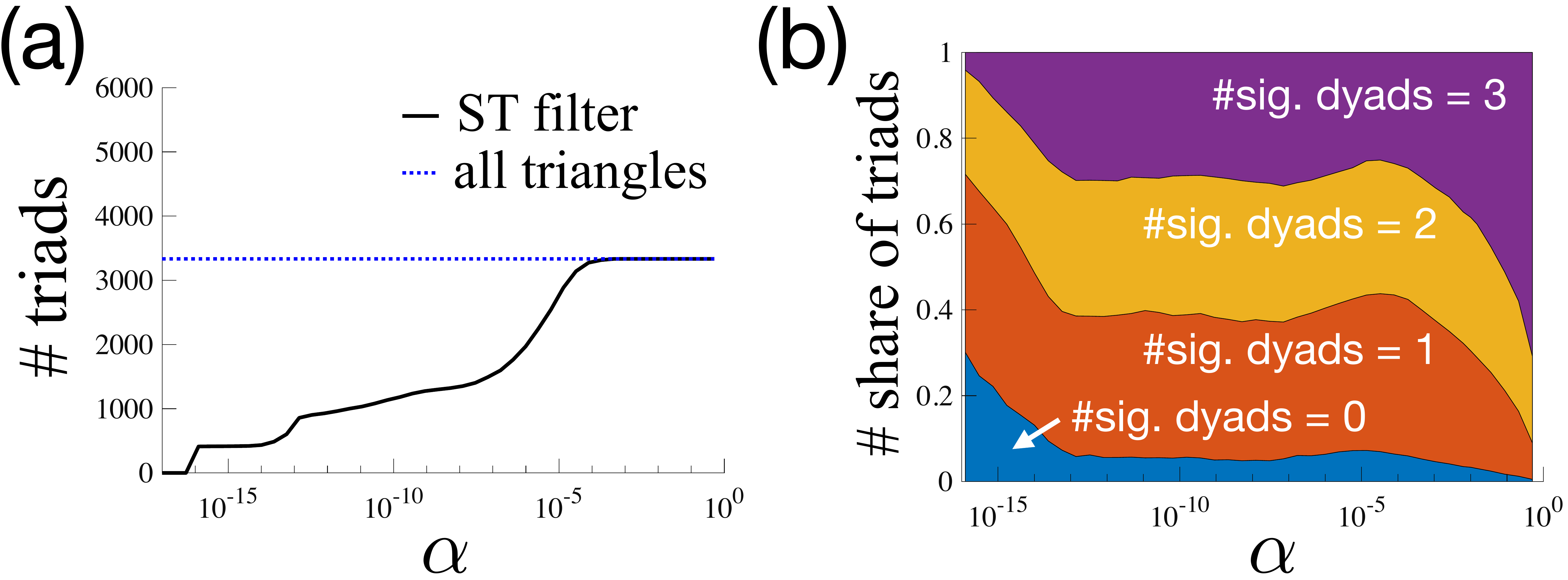}
    \end{center}
    \caption{Significant triadic relationships in the Highschool data set, for a temporal resolution $\Delta=1 {\rm min}$. (a) Number of significant triads
    as a function of $\alpha$.
    (b) Fraction of significant triadic relationships with a given number of significant dyads, as a function of $\alpha$.}
 \label{fig:triangle}
\end{figure}

\subsubsection{Triadic relationships}

As described above, we can \new{use the temporal null model defined
above to} 
extract \new{significant temporal structures. We illustrate this
in the case of}
triads with a significant number of simultaneous interactions.
\new{Note that such a task is by construction impossible in the filtering
methods defined on static aggregated networks, as no temporal constraint
can be detected: for instance, a triangle $(i,j,k)$ in an aggregated
network can in fact result from interactions between the 
pairs of nodes $(i,j)$, $(j,k)$ and $(k,i)$ occurring at different
times.}
\new{Moreover, the null models for the DP and the ECM (and thus the ECM-R) filters are designed to detect significant dyads and we 
cannot directly exploit their null distributions to test the significance of triads or other structures; therefore, the only way to define
a structure as significant is to impose that all the ties 
it includes are significant. 
In the case of the ECM-R, which corresponds to
applying the ECM to each snapshot, it means that 
we can in this case also define the significance of temporal structures,
albeit in a somehow trivial way that is not grounded
in a temporal null model: for instance, we define 
a simultaneous triad as significant if there is at least one
snapshot in which its $3$ edges co-exist and are all $3$ significant.
}

We show in Fig.~\ref{fig:triangle}{\sf a} the number of significant triadic relationships as a function of the significance level $\alpha$, for the Highschool data set and temporal resolution $\Delta = 1 {\rm min}$. Similar results are shown in Fig.~\ref{fig:num_triangle} for the other data sets and different values of $\Delta$. 
For filtering levels $\alpha>10^{-4}$, 
almost all the triangles present in the aggregated network
are considered as significant, \new{except for the Interbank and UK-airline data}.
However, the number of significant triadic relationships  decreases as $\alpha$ becomes lower,
determining a set of triads such that their number of simultaneous interactions cannot be explained by the temporal null model and the
individual nodes' activity levels. 
\new{We show in the SI (Fig.~\ref{fig:num_triangle_ECMR}) 
the number of significant simultaneous triads in the ECM-R
filter: it decreases very fast as the significance level increases,
showing that the ST filter is more able to detect such structures.}

Figure \ref{fig:triangle}{\sf b} highlights moreover a striking feature 
of the significant triads \new{detected by our temporal null model, and
absent by definition in the ECM-R case}, namely, that they 
do not necessarily correspond to three significant ties. In fact, the number of significant ties in a significant triad
can take any value between $0$ and $3$ (see also Fig.~\ref{fig:share_triangle}). 
Reciprocally, not all triangles made by three significant ties
turn out to be significant triads (Fig.~\ref{fig:frac_sigtri}).
This clearly shows how the temporal null model allows us to go beyond the definition of significant ties and find significant higher order structures that could not be unveiled by a static approach. Indeed, considering triangles made by significant ties does not guarantee that the corresponding triads have (a significant number of) simultaneous interactions, while on the other hand ties $(i,j)$ with non-significant number of interactions when considered as dyads can turn out to interact a significant fraction of times \textit{simultaneously with} two other dyads $(j,k)$
and $(i,k)$ for a certain $k$, forming thus a significant triad.

 \section{Discussion} \label{sec:discussion}

In this paper, we have presented a new method to find 
significant ties and structures in temporal network data sets. To this aim,
we have defined a null model of interactions that takes into account
the heterogeneity in the activity of individual nodes and the temporal
dimension of the system, and can potentially be extended to 
include temporal variations in the overall network activity, due
for instance to circadian or weekly rhythms, imposed schedule constraints, etc.
We compute for each pair of nodes the distribution of their number
of interactions in the null model, and compare the empirical value to this
distribution. For any chosen significance level, we thus define as significant pairs
of nodes those with a number of interactions that cannot be explained
by the null model. As the null model includes the heterogeneous activity
of nodes, the temporal network backbone composed by the ties with
significant numbers of interactions is not reducible to the nodes
local properties and contains ties with a broad distribution of numbers
of interactions. Varying the significance level allows us 
to tune the number of node pairs in this backbone. 

\new{We compare the results obtained with our method}
with other backboning methods 
built for weighted static network, hence 
applied here on the temporally aggregated network \new{and with 
a baseline temporal extension of a static filter. This} reveals interesting 
similarities and differences. Our method yields, at 
a given significance level, more pairs of nodes than the other benchmarks.
A more detailed comparison shows that the difference does not come from
ties with large numbers of interactions, which are similarly selected
by all methods, but rather by the fact that our filter uncovers more
significant pairs of nodes with small number of interactions. The resulting
distribution of weights of the significant ties is 
\new{broad, showing the ability of the ST filter to detect significant ties
at all scales.}
Moreover, it turns out that, for networks with a strong
community structure, our method tends to uncover mostly intra-community
ties, showing 
\new{that the weights of} the inter-community ones, once the
node activities are given, \new{can be explained by random
interactions ruled by these activities}.

Thanks to the temporal nature of the null model considered, our method
can also attribute a significance to more complex temporal structures, 
such as sets of simultaneous interactions, which have a clear
 importance in social terms (it is clearly
 not the same to have three interactions $(i,j)$,
 $(j,k)$, $(i,k)$ between three individuals at the same moment
 or at different times), but also for processes
 such as epidemic spread occurring on top of a temporal
 network \cite{gauvin2015revealing}. We have in particular shown that
 significant triads of simultaneous interactions are not equivalent
 to triangles of three significant dyads, illustrating the need to 
 take into account the temporality of these structures, which could
 not be uncovered by static filtering methods.
 
 Our work hints at several perspectives and future research directions.
 First, it would be interesting to refine data representations 
 such as the ones put forward in \cite{Machens:2013,genois2015compensating},
 by combining a backbone at a certain significance level and
 a contact matrix representing in a summarized fashion the non-significant
 ties. Another possibility would be to represent the data as a backbone
 plus the set of activities $\{\{a_i^*\}, \xi(t) \}$. In both cases, these
 representations should be validated by numerical simulations of various 
 types of processes on top of the data. The relevance of such
 representation is two-fold: on the one hand, they allow to
 summarize and generalize complex data sets 
 in a way that can be fed into data-driven models of dynamical processes 
 such as epidemic or information spreading; on the other hand, they keep
 the minimum amount of detailed information on the precise interactions, 
 summarizing less relevant details as distributions or averages, and
 thus in a way that might be easier to render data anonymous. Another
 direction of research would be to define a backbone at a finer resolution,
 namely that would be composed of (sets of) significant interactions instead 
 of ties or set of ties.

\section{Methods}
 
 \new{
\subsubsection*{Data processing} \label{sec:data_process}
 
 Different filtering methods use different network formats. 
 Here we summarize the data processing procedure.
 \begin{itemize}
     \item \textbf{Highschool, Primaryschool, Workplace, Hospital}~\cite{SocioPatterns}\\
     For the ST filter, the network snapshots (i.e., unweighted adjacency matrices) are created so that each represents interactions between people observed over a $\Delta$-minute interval ($\Delta=15$ unless otherwise noted). We exclude the time interval between the last contact of a day and the first contact of the following day. The aggregate network for the DP and ECM filters represents all the interactions recorded over the whole data period, in which edge weights are given by the total numbers of interactions. 
     For the ECM-R filter, we use a sequence of weighted networks aggregated over the timw windows of duration $\Delta$, 
     in which edge weights represent the numbers of interactions observed over a $\Delta$-minute interval.
     
     \item \textbf{Interbank}~\cite{emidHP} \\
     Snapshots for the ST filter are given by daily networks formed by overnight bilateral transactions between banks between June 12, 2007 and July 9, 2007 (20 business days). The daily networks are unweighted and undirected. The edges of the aggregate network for the DP and ECM filters are weighted by the number of transactions observed over the data period. For the ECM-R filter, a weighted networks is created for each day so that the weight of each edge represents the number of intra-day transactions on the corresponding day.
     
     \item \textbf{Email}~\cite{SNAP}\\
     The edges of the network snapshots for the ST filter represent the presence of emails between two members of a research institution in the EU on a given day. The edges of the aggregate network for the DP and ECM filters are weighted by the total numbers of emails over the data period. For the ECM-R filter, we use a sequence of daily weighted networks, in which edge weights represent the numbers of emails exchanged on a given day.

      \item \textbf{LondonBike}~\cite{LondonBike}\\
         The data contains all the trips taken between 00:00 and 23:59 on June 22, 2014 (with time resolution one minute). For the ST filter, each network snapshot represents the trips between stations started within a $15$-minute interval. 
         The aggregate network for the DP and ECM filters represent all the trips recorded over the whole day, in which an edge weight is given by the total number of trips between two stations. For the ECM-R filter, we use a sequence of weighted networks, in which edge weights represent the numbers of trips observed in a given $15$-minute interval. 
     
        \item \textbf{UK-airline}~\cite{UKairline}\\
     Each of the snapshots for the ST filter is an unweighted and undirected adjacency matrix of domestic airlines in the UK in a given year. Edges of the aggregate network for the DP and ECM filters are weighted by the number of snapshots in which the edge is present,
     over the whole data period (1990--2003). For the ECM-R filter, we use as snapshots the yearly weighted networks whose edge weights are given by the number of passengers recorded over the corresponding year, because the number of flights in a given year is not available from the data.

 \end{itemize}
 
 }
 
\subsubsection*{Estimation of nodal activity}\label{sec:activity_estimation}

We perform a maximum likelihood (ML) estimation of $\vect{a}\equiv (a_1,\ldots , a_N)$, taking the $\tau$ temporal snapshots as input, where $\tau = \lfloor T/\Delta\rfloor$. 
If two individuals are independently matched in each time interval according to probability $u(a, a^\prime)$, then the number of times temporal edges are formed between nodes $i$ and $j$ over $\tau$ time intervals 
is a random variable $m_{ij}$ that follows a binomial distribution with parameters $\tau$ and $u(a_{i},a_{j})$. Therefore, the joint probability function leads to
\begin{align}
   p(\{m_{ij}\}|\vect{a}) = 
   \prod_{i,j: i\neq j}\begin{pmatrix} \tau \\ m_{ij}\end{pmatrix} u(a_{i},a_{j})^{m_{ij}} (1-u(a_{i},a_{j}))^{\tau-m_{ij}},
\end{align} 
where $m_{ij}\leq \tau$ denotes the count of temporal edges between $i$ and $j$ observed over $\tau$ periods in the null model. The log-likelihood function 
for the empirical data $\{m_{ij}^{\rm o}\}$ is thus given by
\begin{align}
   \mathcal{L}(\vect{a}) & = \log p(\{m_{ij}^{\rm o}\}|\vect{a}) \notag \\
    &= \sum_{i,j: i\neq j}
    \left[  m_{ij}^{\rm o}\log{(a_{i}a_{j})} + (\tau-m_{ij}^{\rm o}) \log{(1-a_{i}a_{j})}\right]  + \text{const.},
   \label{eq:loglikelihood}
\end{align} 
where ``$\text{const.}$'' denotes the terms that are independent of $\vect{a}$.
\new{Note that the sum runs over all pairs of nodes $(i,j)$, including
those with $\{m_{ij}^{\rm o}\}=0$.}
The ML estimate of $\vect{a}$ is the solution for the following $N$ equations:
\begin{align}
 H_i(\vect{a}^{*}) \equiv & \sum_{j:j\neq i}\frac{m_{ij}^{\rm o}-\tau a_{i}^{*}a_{j}^{*}}{1-a_{i}^{*}a_{j}^{*}} = 0, \; \forall \: i = 1,\ldots, N, \label{eq:ML_a} 
 \end{align}
The first-order condition \eqref{eq:ML_a} is obtained by differentiating the log-likelihood function \eqref{eq:loglikelihood} with respect to $a_i$. The system of $N$ nonlinear equations, $H(\vect{a})=\vect{0}$, can be solved by using a standard numerical algorithm.\footnote{We solve the equation by using Matlab function \url{fsolve}, which is based on a modified Newton method, called the trust-region-dogleg method. The initial values of $a_i$ is given by the configuration model: 
$a_i = \sum_{j:j\neq i}(m_{ij}^{\rm o}/\tau)/\sqrt{2\sum_{i<j}m_{ij}^{\rm o}/\tau}$, where the numerator and the denominator represent the means of $i$'s temporal degree and the doubled number of total temporal edges, respectively.} 
The obtained ML estimates of $\vect{a}$ is denoted by $\vect{a}^*\equiv (a_1^*,\ldots , a_N^*)$. The numbers of contacts obtained from the model and the empirical data are compared in section~\ref{sec:activity_strength}. 
The extension of the method to include time-varying probabilities of creating interactions is shown in section~\ref{sec:SI_timevarying}.

\bigskip

\section*{Acknowledgments}

TK  acknowledges  financial  support  from  the  Japan  Society  for  the  Promotion  of  Science  Grants  no.~15H05729 and 16K03551.

\section*{Author contributions}
All authors designed the study. TK performed the numerical analysis. All authors wrote the manuscript. 

\section*{Competing interests}
The authors declare that they have no conflict of interest.


\begin{thebibliography}{10}
\expandafter\ifx\csname url\endcsname\relax
  \def\url#1{\texttt{#1}}\fi
\expandafter\ifx\csname urlprefix\endcsname\relax\def\urlprefix{URL }\fi
\expandafter\ifx\csname doiprefix\endcsname\relax\def\doiprefix{DOI }\fi
\providecommand{\bibinfo}[2]{#2}
\providecommand{\eprint}[2][]{\url{#2}}

\bibitem{butts2003error}
\bibinfo{author}{Butts, C.~T.}
\newblock \bibinfo{journal}{\bibinfo{title}{Network inference, error, and
  informant (in)accuracy: a Bayesian approach}}.
\newblock {\textit{\JournalTitle{Soc. Netw.}}}
  \textbf{\bibinfo{volume}{25}}, \bibinfo{pages}{103--140}
  (\bibinfo{year}{2003}).

\bibitem{newman2018noisy}
\bibinfo{author}{Newman, M.}
\newblock \bibinfo{journal}{\bibinfo{title}{Network structure from rich but
  noisy data}}.
\newblock {\textit{\JournalTitle{Nature Physics}}} \bibinfo{pages}{in press}
  (\bibinfo{year}{2018}).

\bibitem{newman2018noisyarxiv}
\bibinfo{author}{Newman, M.}
\newblock \bibinfo{journal}{\bibinfo{title}{Network reconstruction and error
  estimation with noisy network data}}.
\newblock {\textit{\JournalTitle{arXiv:1803.02427}}}
  (\bibinfo{year}{2018}).

\bibitem{Cattuto2010PLOS}
\bibinfo{author}{Cattuto, C.} \textit{et~al.}
\newblock \bibinfo{journal}{\bibinfo{title}{Dynamics of person-to-person
  interactions from distributed RFID sensor networks}}.
\newblock {\textit{\JournalTitle{PLOS ONE}}} \textbf{\bibinfo{volume}{5}},
  \bibinfo{pages}{1--9} (\bibinfo{year}{2010}).

\bibitem{Stehle2011PLOS}
\bibinfo{author}{Stehl{\'{e}}, J.} \textit{et~al.}
\newblock \bibinfo{journal}{\bibinfo{title}{{High-resolution measurements of
  face-to-face contact patterns in a primary school.}}}
\newblock {\textit{\JournalTitle{PLOS ONE}}} \textbf{\bibinfo{volume}{6}},
  \bibinfo{pages}{e23176} (\bibinfo{year}{2011}).

\bibitem{JoKarsai2012NewJPhys}
\bibinfo{author}{Jo, H.~H.}, \bibinfo{author}{Karsai, M.},
  \bibinfo{author}{Kertesz, J.} \& \bibinfo{author}{Kaski, K.}
\newblock \bibinfo{journal}{\bibinfo{title}{{Circadian pattern and burstiness
  in mobile phone communication}}}.
\newblock {\textit{\JournalTitle{New J. Phys.}}} \textbf{\bibinfo{volume}{14}},
  \bibinfo{pages}{013055} (\bibinfo{year}{2012}).

\bibitem{Schlapfer2014}
\bibinfo{author}{Schl{\"a}pfer, M.} \textit{et~al.}
\newblock \bibinfo{journal}{\bibinfo{title}{The scaling of human interactions
  with city size}}.
\newblock {\textit{\JournalTitle{Journal of the Royal Society Interface}}}
  \textbf{\bibinfo{volume}{11}}, \bibinfo{pages}{20130789}
  (\bibinfo{year}{2014}).

\bibitem{Centola2010Science}
\bibinfo{author}{Centola, D.}
\newblock \bibinfo{journal}{\bibinfo{title}{{The spread of behavior in an
  online social network experiment}}}.
\newblock {\textit{Science}} \textbf{\bibinfo{volume}{329}},
  \bibinfo{pages}{1194--1197} (\bibinfo{year}{2010}).

\bibitem{sapienza2018NTF}
\bibinfo{author}{Sapienza, A.}, \bibinfo{author}{Bessi, A.} \&
  \bibinfo{author}{Ferrara, E.}
\newblock \bibinfo{journal}{\bibinfo{title}{Non-negative tensor factorization
  for human behavioral pattern mining in online games}}.
\newblock {\textit{Information}} \textbf{\bibinfo{volume}{9}},
  \bibinfo{pages}{66} (\bibinfo{year}{2018}).

\bibitem{Seidman:1983}
\bibinfo{author}{Seidman, S.~B.}
\newblock \bibinfo{journal}{\bibinfo{title}{Network structure and minimum
  degree}}.
\newblock {\textit{\JournalTitle{Soc. Netw.}}} \textbf{\bibinfo{volume}{5}},
  \bibinfo{pages}{269 -- 287} (\bibinfo{year}{1983}).

\bibitem{Alvarez:2005}
\bibinfo{author}{Alvarez-Hamelin, J.~I.}, \bibinfo{author}{Dall'Asta, L.},
  \bibinfo{author}{Barrat, A.} \& \bibinfo{author}{Vespignani, A.}
\newblock \bibinfo{journal}{\bibinfo{title}{K-core decomposition of internet
  graphs: hierarchies, self-similarity and measurement biases}}.
\newblock {\emph{\JournalTitle{Networks and Heterogeneous Media}}}
  \textbf{\bibinfo{volume}{3}}, \bibinfo{pages}{395--411}
  (\bibinfo{year}{2008}).

\bibitem{Kitsak:2010}
\bibinfo{author}{Kitsak, M.} \textit{et~al.}
\newblock \bibinfo{journal}{\bibinfo{title}{Identifying influential spreaders
  in complex networks}}.
\newblock {\textit{\JournalTitle{Nature Physics 6, 888}}}
  (\bibinfo{year}{2010}).

\bibitem{Serrano2009PNAS}
\bibinfo{author}{Serrano, M.~{\'A}.}, \bibinfo{author}{Bogun{\'a}, M.} \&
  \bibinfo{author}{Vespignani, A.}
\newblock \bibinfo{journal}{\bibinfo{title}{Extracting the multiscale backbone
  of complex weighted networks}}.
\newblock {\textit{\JournalTitle{Proceedings of the National Academy of Sciences USA}}} \textbf{\bibinfo{volume}{106}},
  \bibinfo{pages}{6483--6488} (\bibinfo{year}{2009}).

\bibitem{tumminello2011statistically}
\bibinfo{author}{Tumminello, M.}, \bibinfo{author}{Miccich{\`e}, S.},
  \bibinfo{author}{Lillo, F.}, \bibinfo{author}{Piilo, J.} \&
  \bibinfo{author}{Mantegna, R.~N.}
\newblock \bibinfo{journal}{\bibinfo{title}{Statistically validated networks in
  bipartite complex systems}}.
\newblock {\textit{\JournalTitle{PLOS ONE}}} \textbf{\bibinfo{volume}{6}},
  \bibinfo{pages}{e17994} (\bibinfo{year}{2011}).

\bibitem{li2014statistically}
\bibinfo{author}{Li, M.-X.} \textit{et~al.}
\newblock \bibinfo{journal}{\bibinfo{title}{Statistically validated mobile
  communication networks: the evolution of motifs in European and Chinese
  data}}.
\newblock {\textit{\JournalTitle{New Journal of Physics}}}
  \textbf{\bibinfo{volume}{16}}, \bibinfo{pages}{083038}
  (\bibinfo{year}{2014}).

\bibitem{Hatzopoulos2015}
\bibinfo{author}{Hatzopoulos, V.}, \bibinfo{author}{Iori, G.},
  \bibinfo{author}{Mantegna, R.~N.}, \bibinfo{author}{Miccich{\`e}, S.} \&
  \bibinfo{author}{Tumminello, M.}
\newblock \bibinfo{journal}{\bibinfo{title}{Quantifying preferential trading in
  the e-MID interbank market}}.
\newblock {\textit{\JournalTitle{Quantitative Financ.}}}
  \textbf{\bibinfo{volume}{15}}, \bibinfo{pages}{693--710}
  (\bibinfo{year}{2015}).

\bibitem{gemmetto2017arxiv}
\bibinfo{author}{Gemmetto, V.}, \bibinfo{author}{Cardillo, A.} \&
  \bibinfo{author}{Garlaschelli, D.}
\newblock \bibinfo{journal}{\bibinfo{title}{Irreducible network backbones:
  unbiased graph filtering via maximum entropy}}.
\newblock {\textit{\JournalTitle{arXiv:1706.00230}}}
  (\bibinfo{year}{2017}).

\bibitem{casiraghi2017relational}
\bibinfo{author}{Casiraghi, G.}, \bibinfo{author}{Nanumyan, V.},
  \bibinfo{author}{Scholtes, I.} \& \bibinfo{author}{Schweitzer, F.}
\newblock \bibinfo{title}{From relational data to graphs: Inferring significant
  links using generalized hypergeometric ensembles}.
\newblock In \textit{\bibinfo{booktitle}{International Conference on Social
  Informatics}}, \bibinfo{pages}{111--120} (\bibinfo{organization}{Springer},
  \bibinfo{year}{2017}).

\bibitem{marcaccioli2018parametric}
\bibinfo{author}{Marcaccioli, R.} \& \bibinfo{author}{Livan, G.}
\newblock \bibinfo{journal}{\bibinfo{title}{A parametric approach to
  information filtering in complex networks: The \uppercase{P}\'{o}lya
  filter}}.
\newblock {\emph{\JournalTitle{arXiv:1806.09893}}}
  (\bibinfo{year}{2018}).

\bibitem{Holme2012PhysRep}
\bibinfo{author}{Holme, P.} \& \bibinfo{author}{Saram{\"a}ki, J.}
\newblock \bibinfo{journal}{\bibinfo{title}{Temporal networks}}.
\newblock {\textit{\JournalTitle{Phys. Rep.}}}
  \textbf{\bibinfo{volume}{519}}, \bibinfo{pages}{97--125}
  (\bibinfo{year}{2012}).

\bibitem{Masuda2016book}
\bibinfo{author}{Masuda, N.} \& \bibinfo{author}{Lambiotte, R.}
\newblock \textit{\bibinfo{title}{A Guide to Temporal Networks}}
  (\bibinfo{publisher}{World Scientific Publishing}, \bibinfo{year}{2016}).

\bibitem{grabowicz2014fast}
\bibinfo{author}{Grabowicz, P.~A.}, \bibinfo{author}{Aiello, L.~M.} \&
  \bibinfo{author}{Menczer, F.}
\newblock \bibinfo{journal}{\bibinfo{title}{Fast filtering and animation of
  large dynamic networks}}.
\newblock {\textit{\JournalTitle{EPJ Data Science}}}
  \textbf{\bibinfo{volume}{3}}, \bibinfo{pages}{27} (\bibinfo{year}{2014}).
  

\bibitem{Kovanen2011JStatMech}
\bibinfo{author}{Kovanen, L.}, \bibinfo{author}{Karsai, M.},
  \bibinfo{author}{Kaski, K.}, \bibinfo{author}{Kert\'{e}sz, J.} \&
  \bibinfo{author}{Saram\"{a}ki, J.}
\newblock \bibinfo{journal}{\bibinfo{title}{{Temporal motifs in time-dependent
  networks}}}.
\newblock {\textit{\JournalTitle{Journal of Statistical Mechanics}}}
  \bibinfo{pages}{P11005} (\bibinfo{year}{2011}).

\bibitem{Granovetter1973AJS}
\bibinfo{author}{Granovetter, M.}
\newblock \bibinfo{journal}{\bibinfo{title}{{The strength of weak ties}}}.
\newblock {\textit{\JournalTitle{Am. J. Sociol.}}} \textbf{\bibinfo{volume}{78}},
  \bibinfo{pages}{1360--1380} (\bibinfo{year}{1973}).

\bibitem{Watts1998Nature}
\bibinfo{author}{Watts, D.~J.} \& \bibinfo{author}{Strogatz, S.~H.}
\newblock \bibinfo{journal}{\bibinfo{title}{Collective dynamics of
  `small-world' networks}}.
\newblock {\textit{Nature}} \textbf{\bibinfo{volume}{393}},
  \bibinfo{pages}{440--442} (\bibinfo{year}{1998}).

\bibitem{SocioPatterns}
\bibinfo{title}{{\url{http://www.sociopatterns.org/}}}.

\bibitem{Fournet:PLOS2015}
\bibinfo{author}{Mastrandrea, R.}, \bibinfo{author}{Fournet, J.} \&
  \bibinfo{author}{Barrat, A.}
\newblock \bibinfo{journal}{\bibinfo{title}{Contact patterns in a high school:
  A comparison between data collected using wearable sensors, contact diaries
  and friendship surveys}}.
\newblock {\textit{\JournalTitle{PLOS ONE}}} \textbf{\bibinfo{volume}{10}},
  \bibinfo{pages}{1--26} (\bibinfo{year}{2015}).

\bibitem{Genois:2017}
\bibinfo{author}{G\'enois, M.} \& \bibinfo{author}{Barrat, A.}
\newblock \bibinfo{journal}{\bibinfo{title}{Can co-location be used as a proxy
  for face-to-face contacts?}}
\newblock {\textit{\JournalTitle{arXiv:1712.06346}}}  (\bibinfo{year}{2017}).

\bibitem{Vanhems:2013}
\bibinfo{author}{Vanhems, P.} \textit{et~al.}
\newblock \bibinfo{journal}{\bibinfo{title}{Estimating potential infection
  transmission routes in hospital wards using wearable proximity sensors}}.
\newblock {\textit{\JournalTitle{PLOS ONE}}} \textbf{\bibinfo{volume}{8}},
  \bibinfo{pages}{e73970} (\bibinfo{year}{2013}).

\bibitem{kobayashi2018social}
\bibinfo{author}{Kobayashi, T.} \& \bibinfo{author}{Takaguchi, T.}
\newblock \bibinfo{journal}{\bibinfo{title}{Social dynamics of financial
  networks}}.
\newblock {\emph{\JournalTitle{EPJ Data Science}}}
  \textbf{\bibinfo{volume}{7}}, \bibinfo{pages}{15} (\bibinfo{year}{2018}).

\bibitem{kobayashi2018extracting}
\bibinfo{author}{Kobayashi, T.}, \bibinfo{author}{Sapienza, A.} \&
  \bibinfo{author}{Ferrara, E.}
\newblock \bibinfo{journal}{\bibinfo{title}{Extracting the multi-timescale
  activity patterns of online financial markets}}.
\newblock {\emph{\JournalTitle{Sci. Rep.}}}
  \textbf{\bibinfo{volume}{8}}, \bibinfo{pages}{11184} (\bibinfo{year}{2018}).
  
  
\bibitem{emidHP}
\bibinfo{note}{\url{http://www.e-mid.it/}}.

\bibitem{SNAP}
\bibinfo{note}{\url{http://snap.stanford.edu/data/index.html}}.

\bibitem{LondonBike}
\bibinfo{author}{Munoz-Mendez, F.}, \bibinfo{author}{Klemmer, K.},
  \bibinfo{author}{Han, K.} \& \bibinfo{author}{Jarvis, S.}
\newblock \bibinfo{journal}{\bibinfo{title}{Community structures, interactions
  and dynamics in london's bicycle sharing network}}.
\newblock {\emph{\JournalTitle{arXiv:1804.05584}}}


\bibitem{UKairline}
\bibinfo{author}{Morer, I.}, \bibinfo{author}{Cardillo, A.},
  \bibinfo{author}{Diaz-Guilera, A.}, \bibinfo{author}{Prignano, L.} \&
  \bibinfo{author}{Lozano, S.}
\newblock \bibinfo{journal}{\bibinfo{title}{Comparing spatial networks: A `one
  size fits all' efficiency-driven approach}}.
\newblock {\emph{\JournalTitle{arXiv:1807.00565}}}
  (\bibinfo{year}{2018}).

\bibitem{Rosvall2008PNAS}
\bibinfo{author}{Rosvall, M.} \& \bibinfo{author}{Bergstrom, C.~T.}
\newblock \bibinfo{journal}{\bibinfo{title}{Maps of random walks on complex
  networks reveal community structure}}.
\newblock {\emph{\JournalTitle{Proc. Natl. Acad. Sci. USA}}}
  \textbf{\bibinfo{volume}{105}}, \bibinfo{pages}{1118--1123}
  (\bibinfo{year}{2008}).

\bibitem{HolmeSaramaki2013book_Springer}
\bibinfo{author}{Holme, P.} \& \bibinfo{author}{Saram\"{a}ki, J.}
\newblock \textit{\bibinfo{title}{{Temporal Networks}}}
  (\bibinfo{publisher}{Springer-Verlag}, \bibinfo{address}{Berlin},
  \bibinfo{year}{2013}).

\bibitem{kobayashi2017significant}
\bibinfo{author}{Kobayashi, T.} \& \bibinfo{author}{Takaguchi, T.}
\newblock \bibinfo{journal}{\bibinfo{title}{Identifying
  relationship lending in the interbank market: A network approach}}.
\newblock {\textit{\JournalTitle{arXiv:1708.08594}}}
  (\bibinfo{year}{2017}).

\bibitem{Caldarelli2002PhysRevLett}
\bibinfo{author}{Caldarelli, G.}, \bibinfo{author}{Capocci, A.},
  \bibinfo{author}{De~Los~Rios, P.} \& \bibinfo{author}{Mu\~{n}oz, M.~A.}
\newblock \bibinfo{journal}{\bibinfo{title}{{Scale-free networks from varying
  vertex intrinsic fitness}}}.
\newblock {\textit{\JournalTitle{Phys. Rev. Lett.}}}
  \textbf{\bibinfo{volume}{89}}, \bibinfo{pages}{258702}
  (\bibinfo{year}{2002}).

\bibitem{Boguna2003PhysRevE}
\bibinfo{author}{Bogu\~n\'a, M.} \& \bibinfo{author}{Pastor-Satorras, R.}
\newblock \bibinfo{journal}{\bibinfo{title}{{Class of correlated random
  networks with hidden variables}}}.
\newblock {\textit{\JournalTitle{Phys. Rev. E}}} \textbf{\bibinfo{volume}{68}},
  \bibinfo{pages}{036112} (\bibinfo{year}{2003}).

\bibitem{DeMasi2006PRE}
\bibinfo{author}{De~Masi, G.}, \bibinfo{author}{Iori, G.} \&
  \bibinfo{author}{Caldarelli, G.}
\newblock \bibinfo{journal}{\bibinfo{title}{Fitness model for the Italian
  interbank money market}}.
\newblock {\textit{\JournalTitle{Phys. Rev. E}}} \textbf{\bibinfo{volume}{74}},
  \bibinfo{pages}{066112} (\bibinfo{year}{2006}).

\bibitem{STcode}
\bibinfo{note}{\url{http://doi.org/10.5281/zenodo.1243994 }}.

\bibitem{Newman_Girvan2004PRE}
\bibinfo{author}{Newman, M. E.~J.} \& \bibinfo{author}{Girvan, M.}
\newblock \bibinfo{journal}{\bibinfo{title}{Finding and evaluating community
  structure in networks}}.
\newblock {\textit{\JournalTitle{Phys. Rev. E}}} \textbf{\bibinfo{volume}{69}},
  \bibinfo{pages}{026113} (\bibinfo{year}{2004}).

\bibitem{newman2004analysis}
\bibinfo{author}{Newman, M.~E.}
\newblock \bibinfo{journal}{\bibinfo{title}{Analysis of weighted networks}}.
\newblock {\textit{\JournalTitle{Physical Rev. E}}}
  \textbf{\bibinfo{volume}{70}}, \bibinfo{pages}{056131}
  (\bibinfo{year}{2004}).

\bibitem{Onnela2007PNAS}
\bibinfo{author}{Onnela, J.-P.} \textit{et~al.}
\newblock \bibinfo{journal}{\bibinfo{title}{{Structure and tie strengths in
  mobile communication networks.}}}
\newblock {\textit{\JournalTitle{Proc. Natl. Acad. Sci. USA}}}
  \textbf{\bibinfo{volume}{104}}, \bibinfo{pages}{7332--7336}
  (\bibinfo{year}{2007}).

\bibitem{Machens:2013}
\bibinfo{author}{Machens, A.} \textit{et~al.}
\newblock \bibinfo{journal}{\bibinfo{title}{An infectious disease model on
  empirical networks of human contact: bridging the gap between dynamic network
  data and contact matrices}}.
\newblock {\textit{\JournalTitle{BMC Infectious Diseases}}}
  \textbf{\bibinfo{volume}{13}}, \bibinfo{pages}{185} (\bibinfo{year}{2013}).

\bibitem{genois2015compensating}
\bibinfo{author}{G{\'e}nois, M.}, \bibinfo{author}{Vestergaard, C.~L.},
  \bibinfo{author}{Cattuto, C.} \& \bibinfo{author}{Barrat, A.}
\newblock \bibinfo{journal}{\bibinfo{title}{Compensating for population
  sampling in simulations of epidemic spread on temporal contact networks}}.
\newblock {\textit{\JournalTitle{Nature Commun.}}}
  \textbf{\bibinfo{volume}{6}} (\bibinfo{year}{2015}).

\bibitem{gauvin2015revealing}
\bibinfo{author}{Gauvin, L.}, \bibinfo{author}{Panisson, A.},
  \bibinfo{author}{Barrat, A.} \& \bibinfo{author}{Cattuto, C.}
\newblock \bibinfo{journal}{\bibinfo{title}{Revealing latent factors of
  temporal networks for mesoscale intervention in epidemic spread}}.
\newblock {\textit{\JournalTitle{arXiv:1501.02758}}}
  (\bibinfo{year}{2015}).

\bibitem{LeCam1960}
\bibinfo{author}{Le~Cam, L.}
\newblock \bibinfo{journal}{\bibinfo{title}{An approximation theorem for the
  Poisson binomial distribution}}.
\newblock {\textit{\JournalTitle{Pacific Journal of Mathematics}}}
  \textbf{\bibinfo{volume}{10}}, \bibinfo{pages}{1181--1197}
  (\bibinfo{year}{1960}).

\bibitem{Barbour1983poisson}
\bibinfo{author}{Barbour, A.} \& \bibinfo{author}{Eagleson, G.}
\newblock \bibinfo{journal}{\bibinfo{title}{Poisson approximation for some
  statistics based on exchangeable trials}}.
\newblock {\textit{\JournalTitle{Advances in Applied Probability}}}
  \textbf{\bibinfo{volume}{15}}, \bibinfo{pages}{585--600}
  (\bibinfo{year}{1983}).

\bibitem{Steele1994LeCam}
\bibinfo{author}{Steele, J.~M.}
\newblock \bibinfo{journal}{\bibinfo{title}{Le cam's inequality and Poisson
  approximations}}.
\newblock {\textit{\JournalTitle{The American Mathematical Monthly}}}
  \textbf{\bibinfo{volume}{101}}, \bibinfo{pages}{48--54}
  (\bibinfo{year}{1994}).

\bibitem{squartini2011analytical}
\bibinfo{author}{Squartini, T.} \& \bibinfo{author}{Garlaschelli, D.}
\newblock \bibinfo{journal}{\bibinfo{title}{Analytical maximum-likelihood
  method to detect patterns in real networks}}.
\newblock {\textit{\JournalTitle{New Journal of Physics}}}
  \textbf{\bibinfo{volume}{13}}, \bibinfo{pages}{083001}
  (\bibinfo{year}{2011}).

\bibitem{mastrandrea2014enhanced}
\bibinfo{author}{Mastrandrea, R.}, \bibinfo{author}{Squartini, T.},
  \bibinfo{author}{Fagiolo, G.} \& \bibinfo{author}{Garlaschelli, D.}
\newblock \bibinfo{journal}{\bibinfo{title}{Enhanced reconstruction of weighted
  networks from strengths and degrees}}.
\newblock {\textit{\JournalTitle{New Journal of Physics}}}
  \textbf{\bibinfo{volume}{16}}, \bibinfo{pages}{043022}
  (\bibinfo{year}{2014}).

\bibitem{squartini2015unbiased}
\bibinfo{author}{Squartini, T.}, \bibinfo{author}{Mastrandrea, R.} \&
  \bibinfo{author}{Garlaschelli, D.}
\newblock \bibinfo{journal}{\bibinfo{title}{Unbiased sampling of network
  ensembles}}.
\newblock {\textit{\JournalTitle{New Journal of Physics}}}
  \textbf{\bibinfo{volume}{17}}, \bibinfo{pages}{023052}
  (\bibinfo{year}{2015}).

\bibitem{MaxSamcode}
\bibinfo{note}{\url{https://jp.mathworks.com/matlabcentral/fileexchange/46912-max-sam-package-zip}}.

\bibitem{Zhao2011PRE}
\bibinfo{author}{Zhao, K.}, \bibinfo{author}{Stehl{\'{e}}, J.},
  \bibinfo{author}{Bianconi, G.} \& \bibinfo{author}{Barrat, A.}
\newblock \bibinfo{journal}{\bibinfo{title}{{Social network dynamics of
  face-to-face interactions}}}.
\newblock {\emph{\JournalTitle{Physical Review E}}}
  \textbf{\bibinfo{volume}{83}}, \bibinfo{pages}{056109} (\bibinfo{year}{2011}).
  
\end{thebibliography}

\newpage

\setcounter{section}{0}
\setcounter{table}{0}
\setcounter{equation}{0}
\setcounter{figure}{0}
\setcounter{page}{1}
     
\renewcommand{\thetable}{S\arabic{table}}
\renewcommand{\thefigure}{S\arabic{figure}}
\renewcommand{\thesection}{S\arabic{section}}
\renewcommand{\theequation}{S\arabic{equation}}



\begin{center}
{\Large{\textbf{Supplementary Information \\ \bigskip
``The structured backbone of temporal social ties"}}}\\ \bigskip
\large{Teruyoshi Kobayashi, Taro Takaguchi, Alain Barrat}

\end{center}

\begin{figure}[thb]
     \centering
     \includegraphics[width=.75\columnwidth]{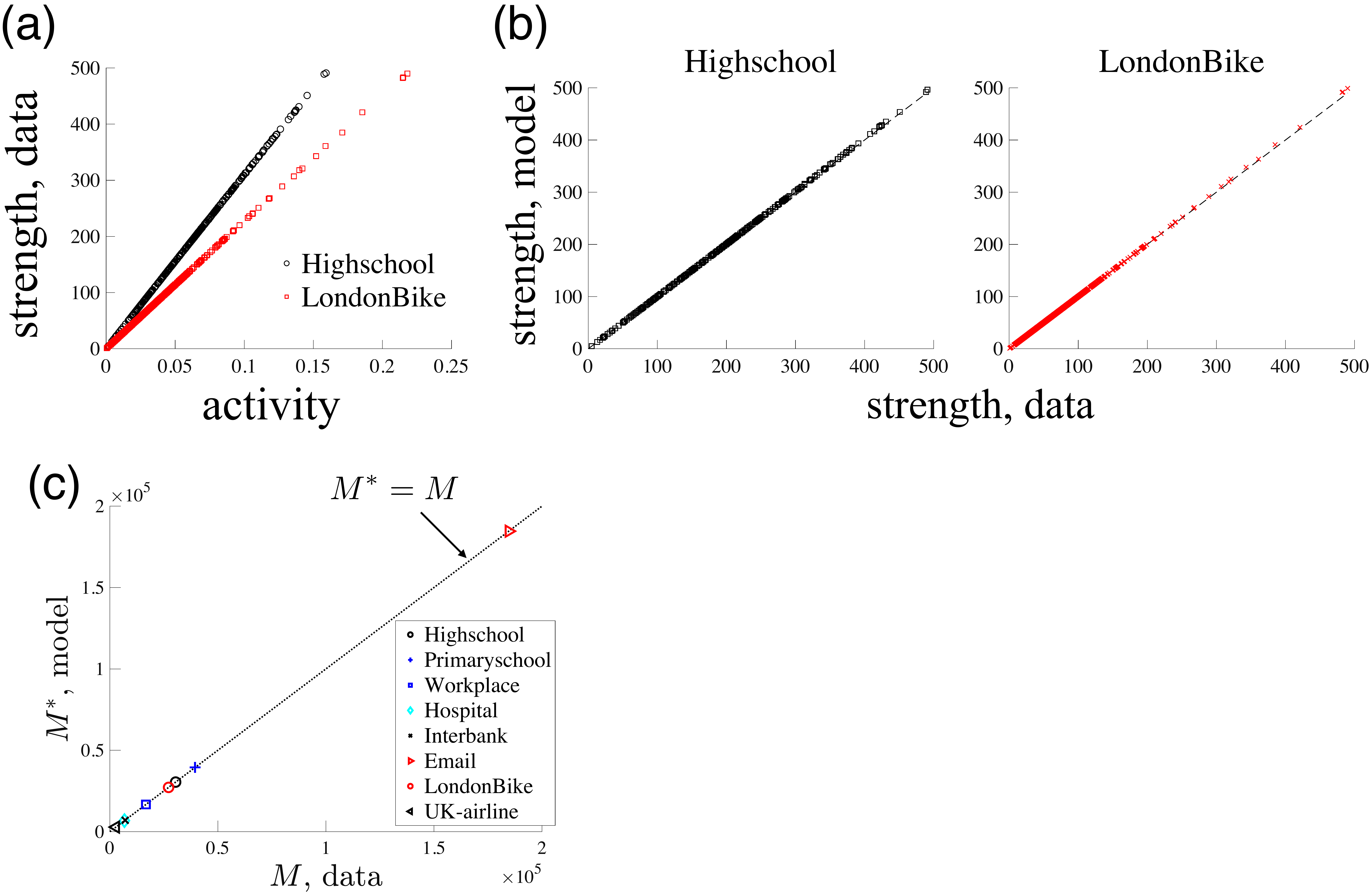}
     \caption{\new{Comparison between the model and the data. (a) Empirical strength vs. estimated activity. Each dot represents a node. Activity is estimated using the temporal null model described in the main text (Eq.~\ref{eq:matching_func}). The strength denotes the total number of snapshot edges emanating from a node. (b) Theoretical vs. empirical strength. The dashed line denotes 45-degree line. The theoretical strength of node $i$ with estimated activity $a_i^*$ is given by $\tau\sum_{j\neq i}u(a_i^*,a_j^*)$. (c) Comparison between the total number of snapshot edges in the data 
     and the model.}}
     \label{fig:activity_strength}
 \end{figure}

\new{
\section{\new{Relationship between activity, strength and degree}}
\label{sec:activity_strength}

 \subsection{Model fit}
 In the temporal null model, the parameters $\{a_i\}$ represent
 the intrinsic activities of nodes, which then
 determine their probability of interactions with other nodes. 
 When considering a data set, these activities are thus not 
 directly observable, but can be 
  estimated by a maximum likelihood method as described in the Methods
  section. 
  Given its definition, the estimated activity of a node in a temporal network is expected to be related to the total number of snapshot 
  edges involving that node, namely its strength. Figure~\ref{fig:activity_strength}a shows indeed that 
   the estimated activity is proportional to the empirical strength.
   In Fig. \ref{fig:activity_strength}b  we also show that the empirical strength is correctly predicted by 
    the value obtained in the model, namely $\tau\sum_{j\neq i}u(a_i^*,a_j^*)$. 
 Figure~\ref{fig:activity_strength}c finally compares the total numbers of snapshot edges in the data ($M=\sum_{i<j} m_{ij}$) and the model 
 $M^* = \tau\sum_{i<j}u(a_i^*,a_j^*)$. 
 The almost perfect fit between the estimated and empirical values of the strength and of the total numbers of edges 
 indicates that the maximum likelihood estimation of the activity
 vector works well for a wide range of temporal-network data.
 
}

\begin{figure}[thb]
\begin{center}
\includegraphics[width=.8\columnwidth]{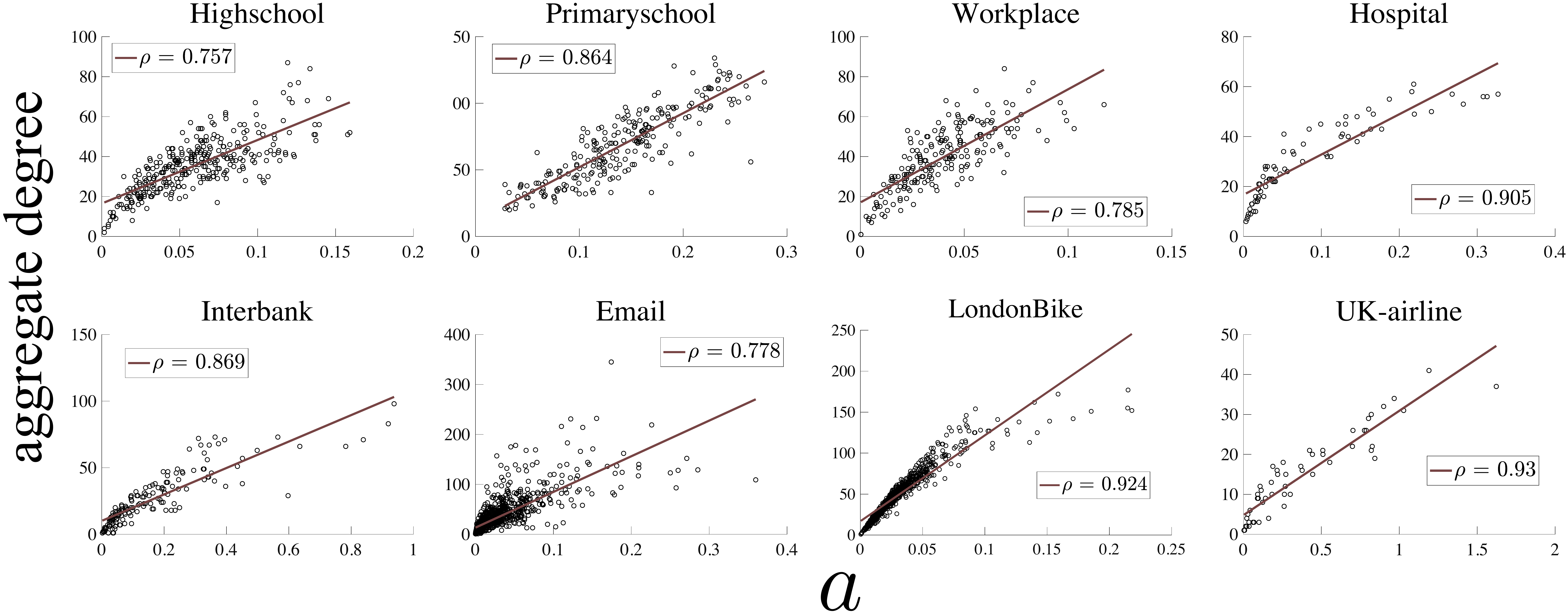}
    \end{center}
    \caption{\new{Aggregate degree vs. activity. Activities are estimated using the temporal null model (Eq.~\ref{eq:matching_func}). $\rho$ denotes the Pearson correlation coefficient. 
    The aggregate degree (i.e., the number of distinct neighbors) has a positive correlation with the estimated activity.}}
 \label{fig:degree_activity}
\end{figure}

\begin{figure}[thb]
\begin{center}
\includegraphics[width=.8\columnwidth]{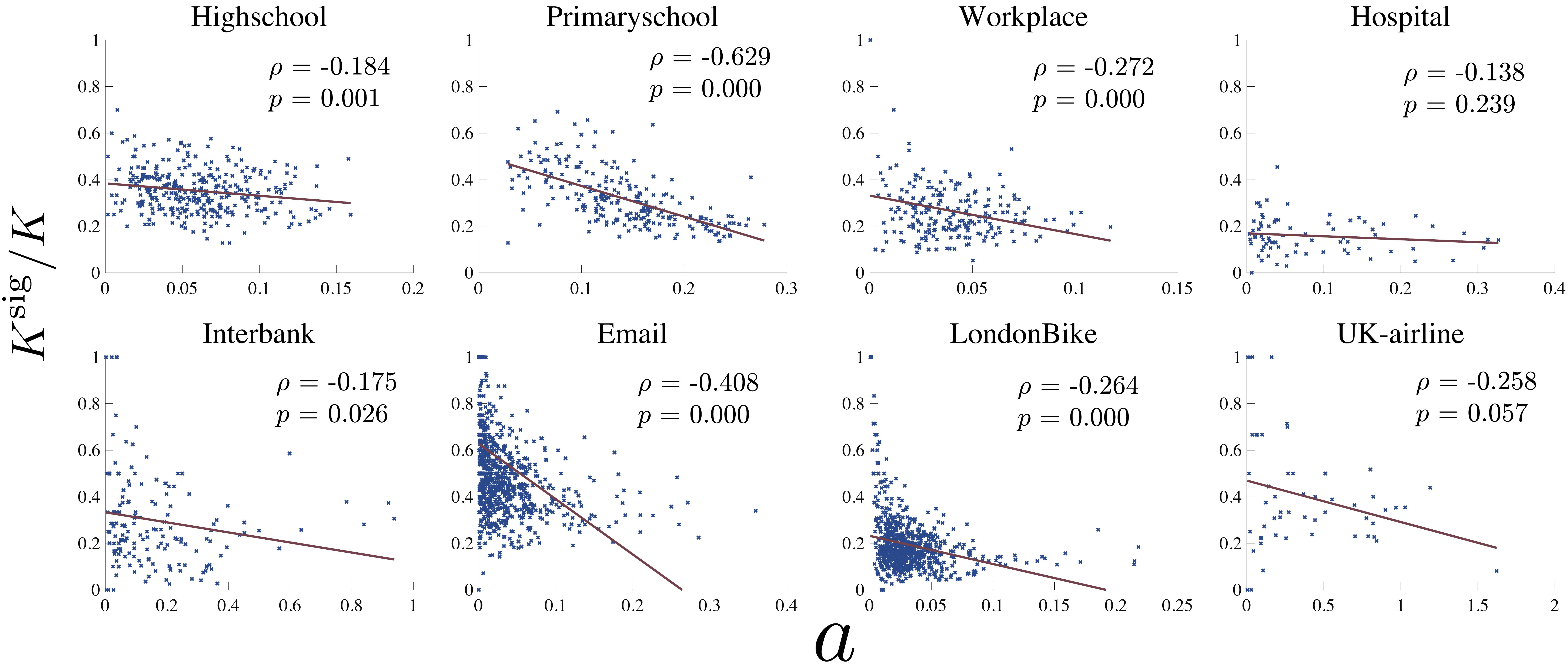}
    \end{center}
    \caption{\new{Negative correlation between the share of significant ties and activity. $K^{\rm sig}$ denotes the number of significant ties of a node (here at $\alpha=0.01$) while $K$ is its aggregate degree. Each dot denotes a node, and the solid line shows a linear regression. $\rho$ and $p$ denote the Pearson correlation coefficient and its $p$-value, respectively. Nodes with lower activities are 
    slightly more likely to have significant ties. 
    }}
 \label{fig:activity_corr}
\end{figure}

 \new{
\subsection{Node degree and significant ties}
Our temporal null model does not consider the  
node aggregate degree as an argument of the interaction probability $u$
(the aggregate degree is the number of 
  neighbors in the aggregate network, i.e., number of distinct 
  nodes with whom a node has had at least one interaction). 
A first motivation for this is to avoid dealing with a more complex model
and a large number of additional parameters. 

For completeness, we examine here the correlations between activity, degree
and significant ties. First, Fig. \ref{fig:degree_activity} 
shows that the degree has a strong positive
correlation with the estimated activity in all data sets.
This suggests that the estimated activities carry some information
on the aggregate degree.

We also show in Fig.~\ref{fig:activity_corr} that there
actually exists a weak negative correlation between activity and the fraction of significant edges emanating from a node, represented by $K^{\rm sig}/K$, where $K^{\rm sig}$ denotes the number of significant edges. This would tend
to show the relevance of enriching the null model with degree information.
However, we do not find any significant correlation between 
$K$ and $K^{\rm sig}/K$ \textit{for a given level of activity}. 
This indicates that once activity is taken into account, and as
it is already strongly correlated with the degree, 
node degree would not be more informative in predicting 
the likelihood of having significant ties. 
}

\section{Time-varying matching probability} \label{sec:SI_timevarying}

In constructing the temporal fitness model, it is possible to 
take into account the possibility that the probability of an interaction between 
two nodes can vary over time even when individuals' intrinsic activities $\{a_i\}$ are constant. 
This can happen when, for example, a school schedule has a certain rhythm (e.g., lunch time, class schedule, etc), or due to circadian or weekly rhythms. 
The probability $u$ for the existence of an interaction between two nodes
$i$ and $j$ at time $t$ is then given as
\begin{align}
    u(a_i,a_j,t) \equiv a_ia_j\xi(t), \;\; t=1,\ldots , \tau,
\end{align}
where $\xi(t)$ denotes a time-varying parameter. We assume that there is no 
correlation between the values of $\xi$
at different times, and the interaction probabilities are independent across time intervals.

 The joint probability function for a certain temporal network $\{A_{t}\}$
is obtained as
\begin{align}
   p(\{{A}_{t}\}|\vect{a},\vect{\xi}) = \prod_{t=1}^{\tau}\prod_{i,j: i\neq j} u(a_{i},a_{j},t)^{A_{ij,t}} (1-u(a_{i},a_{j},t))^{1-A_{ij,t}},
\end{align} 
where $A_{ij,t}$ is the $(i,j)$ element of the adjacency matrix in time interval $t$, denoted by ${A}_t$, and $\vect{\xi}\equiv (\xi(1),\ldots,\xi(\tau))^\top$. 
 The log-likelihood function is thus given by
\begin{align}
   \mathcal{L}(\vect{a},\vect{\xi}) & = \log p(\{{A}_{t}\}|\vect{a},\vect{\xi}) \notag \\
    &= \sum_{t=1}^{\tau}\sum_{i,j: i\neq j}\left[  A_{ij,t}\log{(a_{i}a_{j}\xi(t))} + (1-A_{ij,t}) \log{(1-a_{i}a_{j}\xi(t))}\right],
   \label{eq:loglikelihood_var}
\end{align} 
The maximum-likelihood estimate of $(\vect{a},\vect{\xi})$ is the solution for the following $N+\tau-1$ equations:
\begin{align}
 H_i^{\rm act}(\vect{a}^{*},\vect{\xi}^{*}) \equiv & \sum_{t=1}^{\tau}\sum_{j:j\neq i}\frac{A_{ij,t}- a_{i}^{*}a_{j}^{*}\xi^{*}(t)}{1-a_{i}^{*}a_{j}^{*}\xi^{*}(t)} = 0, \;  \: i = 1,\ldots, N, \label{eq:ML_av}\\
 H_t^{\rm time}(\vect{a}^{*},\vect{\xi}^{*}) \equiv & \sum_{i,j:j\neq i}\frac{A_{ij,t}- a_{i}^{*}a_{j}^{*}\xi^{*}(t)}{1-a_{i}^{*}a_{j}^{*}\xi^{*}(t)} = 0, \;  \: t = 2,\ldots, \tau, 
 \label{eq:ML_xi} 
 \end{align}
The first-order conditions \eqref{eq:ML_av} and \eqref{eq:ML_xi} are obtained by differentiating the log-likelihood function Eq.~\eqref{eq:loglikelihood_var} with respect to $a_i$ for $i=1,\ldots N$ and $\xi(t)$ for $t=2,\ldots , \tau$. 
For $t=1$, $\xi(1)$ is normalized as one since otherwise there would arise a linear dependency between the optimality conditions and therefore the solution would be indeterminate. This reflects the fact that any combination of $\hat{a}_i$, $\hat{a}_j$ and ${\hat\xi(t)}$ would satisfy the optimality conditions if $a_i^*a_j^* = c\cdot \hat{a}_i\hat{a}_j$ and $\xi^*(t)=\hat{\xi}(t)/c$. In solving the nonlinear equations \eqref{eq:ML_av} and \eqref{eq:ML_xi}, the initial values for $a_i$ and $\xi(t)$ are set as $a_i = \sum_{j:j\neq i}(m_{ij}/\tau)/\sqrt{2\sum_{i<j}m_{ij}/\tau}$ and $0.999$, respectively.

 Under the null model with a time-varying term, the average number of contacts between $i$ and $j$ over $\tau$ periods is given by
    \begin{align}
    \lambda_{ij}\equiv\sum_{t=1}^{\tau}u(a_i,a_j,t), \; \forall \; i,j.
    \end{align}
    Thus, the number of contacts obeys a Poisson binomial distribution with mean $\lambda_{ij}$ and variance $\sigma_{ij}\equiv\sum_{t=1}^{\tau}(1-u(a_i,a_j,t))u(a_i,a_j,t)$.
     Since an exact functional form for a Poisson binomial distribution is intractable, we approximate the distribution of $\{m_{ij}\}$ with a Poisson distribution~\cite{LeCam1960,Barbour1983poisson,Steele1994LeCam}:
    \begin{align}
   f(m_{ij}|\vect{a},\vect{\xi}) \approx \frac{\lambda_{ij}^{m_{ij}}e^{-{\lambda_{ij}}}}{m_{ij}!} \equiv \widetilde{f}(m_{ij}|\vect{a},\vect{\xi}), \label{eq:poisson}
   \end{align} 
   where the error bound is given by Le Cam's theorem~\cite{LeCam1960,Barbour1983poisson,Steele1994LeCam}:
   \begin{align}
       \sum_{m_{ij}=0}^\infty\left| f(m_{ij}|\vect{a},\vect{\xi}) - \frac{\lambda_i^{m_{ij}}e^{-\lambda_{ij}}}{m_{ij}!}\right| < \frac{2(1-e^{-\lambda_{ij}})}{\lambda_{ij}}\sum_{t=0}^{\tau}u(a_i,a_j,t)^2,  \; \forall \; i,j.
   \end{align}
   We use Eq.~\eqref{eq:poisson} in testing the significance of edge $(i,j)$ for a given observation $m_{ij}^{\rm o}$.

\begin{figure}[thb]
\begin{center}
           \includegraphics[width=.8\columnwidth]{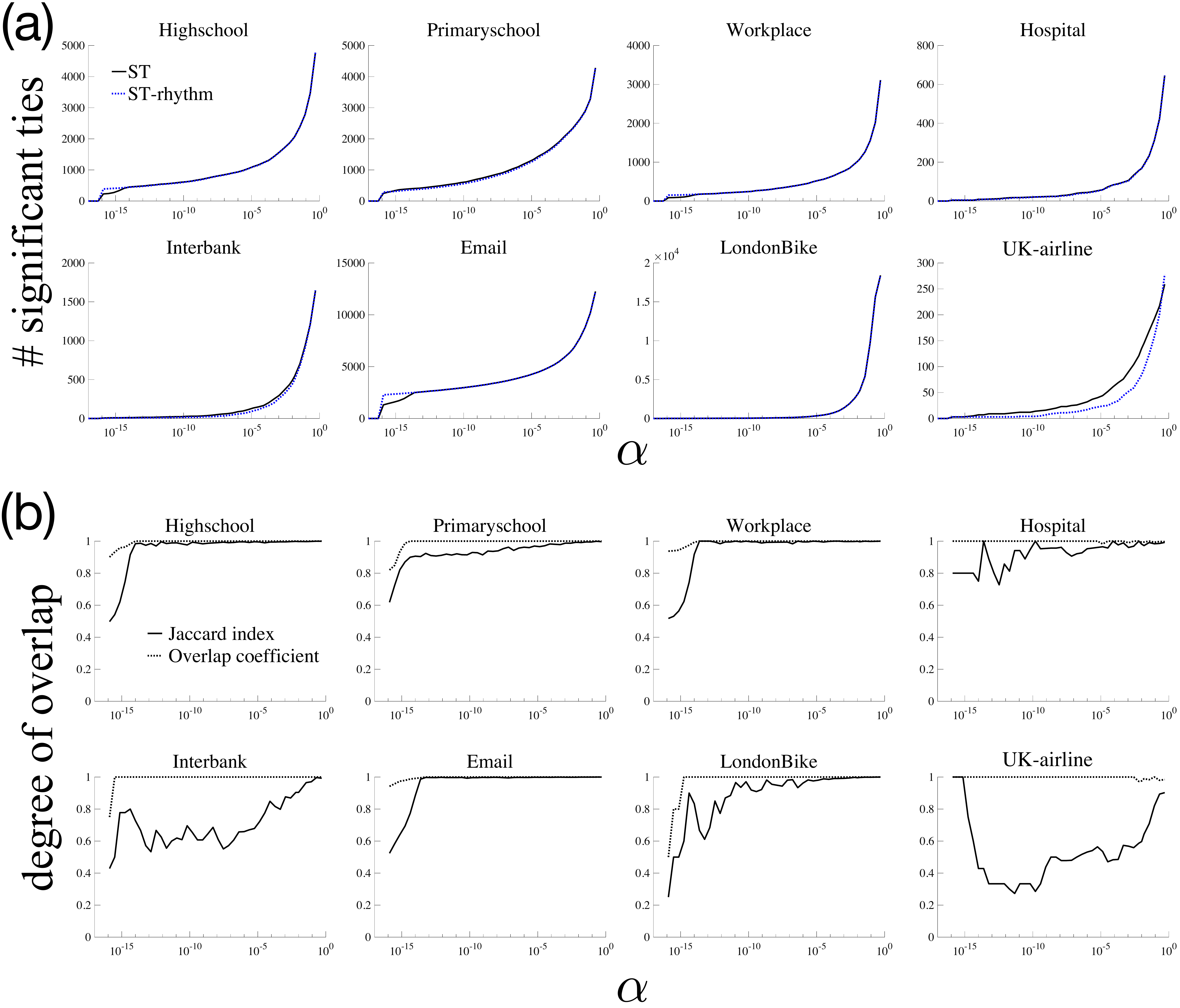}
    \end{center}
    \caption{Comparison between constant and time-varying interaction probabilities. (a) Number of significant ties vs. significance level $\alpha$. \new{ST-rhythm denotes the ST filter with a time-varying parameter.} 
    (b) Jaccard index and overlap coefficient for quantifying the overlap between the lists of 
    significant ties detected by the two null models, vs. $\alpha$. \new{The overlap coefficient, or Szymkiewicz–Simpson coefficient, is defined by $S(I_{\rm ST},I_{\rm ST\text{-}rhythm}) \equiv | I_{\rm ST} \cap I_{\rm ST\text{-}rhythm}|/ \min(| I_{\rm ST}| , |I_{\rm ST\text{-}rhythm}|)$. The fact that it remains equal or very close
    to $1$ indicates that, even in the few cases where the Jaccard index is not very large, the set of significant ties identified by
    the ST-rhythm is included in the set identified by the ST filter.}
    }
 \label{fig:timevarying_prob}
\end{figure}

 A comparison between the models with and without time-varying parameter is shown in Fig.~\ref{fig:timevarying_prob}. It shows that the test results are almost identical between the two null models for
 most data sets; the numbers and the degree of overlap of identified significant ties suggests that the introduction of a time-varying parameter for capturing an activity rhythm does not affect the results shown in the main text.   

\clearpage
\newpage

\section{Backboning methods for static networks} \label{sec:SI_DP_ECM}

We recall here two well-known ways to assign a significance to edges and build backbones 
for static weighted networks. In our context, it can be done
by first aggregating the temporal network on the available time-window, obtaining
a network where the degree of a node is given by its number of distinct neighbors, 
the weight of an edge is total interaction time between two nodes, and the strength
of an edge is given by  $s_i=\sum_{j,t}A_{ij,t}$. 

\subsection{Disparity filter}

 The Disparity (DP) filter~\cite{Serrano2009PNAS} is a filtering algorithm to classify the edges of a static weighted network into significant and insignificant ones. The DP filter uses only local information: the weight of an edge, $\omega_{ij}$, the nodal degree, $k_i$, and the strength, $s_i$. The idea is that if node $i$ has no specific relationship with its neighbors, then its strength (i.e., sum of weights) is distributed uniformly at random on 
 the $k_i$ edges incident to it. 
 The authors of \cite{Serrano2009PNAS} show that the link between $i$ and $j$ is regarded as significant at filtering level $\alpha$, if it satisfies the following condition: 
\begin{align}
    1- (k_i-1)\int_0^{p_{ij}}(1-x)^{k_i-2}dx < \alpha, \label{eq:DPfilter}
\end{align}
where $p_{ij}= \omega_{ij}/s_i$. The LHS of Eq.~\eqref{eq:DPfilter} represents the $p$-value for the null hypothesis that the edge weights are distributed uniformly at random. In fact, as argued in Gemmetto et al.~\cite{gemmetto2017arxiv}, the significance of edge between $i$ and $j$ is not necessarily identical to that between $j$ and $i$ even for an undirected network. Therefore, one needs to test the significance of ``two edges" $(i,j)$ and $(j,i)$ independently, and then the (undirected) edge is regarded as significant if at least one of the two ``edges" satisfies the criterion \eqref{eq:DPfilter}.

\subsection{ECM filter}

The ECM (enhanced configuration model) filter~\cite{gemmetto2017arxiv} is developed based on the idea that statistically significant edges are the ones whose presence cannot be explained by random chance. More specifically, the entropy-maximizing random matching probabilities are calculated with the ECM in which edge weights are distributed at random as uniformly as possible subject to two constraints: $\langle \vec{k}\rangle = \vec{k}^*$ and $\langle \vec{s}\rangle = \vec{s}^*$, where $x^*$ denotes the empirical value of variable $x$. Gemmetto et al.~\cite{gemmetto2017arxiv} show that the $p$-value for edge $(i,j)$ is then given by 
\begin{align}
    \gamma_{ij}^{*} = p_{ij}^*(y_i^{*}y_j^{*})^{w_{ij}^*-1},
\end{align}
where
\begin{align}
    p_{ij}^* = \frac{x_{i}^*x_{j}^*y_{i}^*y_{j}^*}{1-y_{i}^*y_{j}^* + x_{i}^*x_{j}^*y_{i}^*y_{j}^*},
\end{align}
and $x_i^*$ and $y_i^*$ represent \textit{hidden variables} (or auxiliary variables)~\cite{squartini2011analytical,mastrandrea2014enhanced} that solve the following conditions:
\begin{align}
    k_i^* &= \sum_{j\neq i}\frac{x_{i}x_{j}y_{i}y_{j}}{1-y_{i}y_{j} + x_{i}x_{j}y_{i}y_{j}} \;\;\; \forall i, \label{eq:kstar}\\
    s_i^* &= \sum_{j\neq i}\frac{x_{i}x_{j}y_{i}y_{j}}{(1-y_{i}y_{j})(1-y_{i}y_{j} + x_{i}x_{j}y_{i}y_{j})} \;\;\; \forall i.\label{eq:sstar} 
\end{align}
One needs to solve a system of $2N$ nonlinear equations to obtain $\vec{x^*}$ and $\vec{y^*}$. In fact, $-\ln{x}_i$ ($-\ln{y}_i$) corresponds to a Lagrange multiplier associated with the constraint $\langle {k_i}\rangle = k_i^*$ ($\langle {s_i}\rangle = s_i^*$). The backbone of a weighted network with significance level $\alpha$ is the network consisting only of edges $(i,j)\in\{(i,j): \gamma_{ij}^*< \alpha \}$.
Our implementation for the calculation of Eqs.~\eqref{eq:kstar} and \eqref{eq:sstar} is based on the ``Max \& Sam" method proposed in \cite{squartini2015unbiased}, and the MATLAB code is available from \cite{MaxSamcode}.

\section{\new{Generating synthetic temporal networks}} \label{sec:synthetic_process}
\begin{figure}[thb]
\begin{center}
           \includegraphics[width=.8\columnwidth]{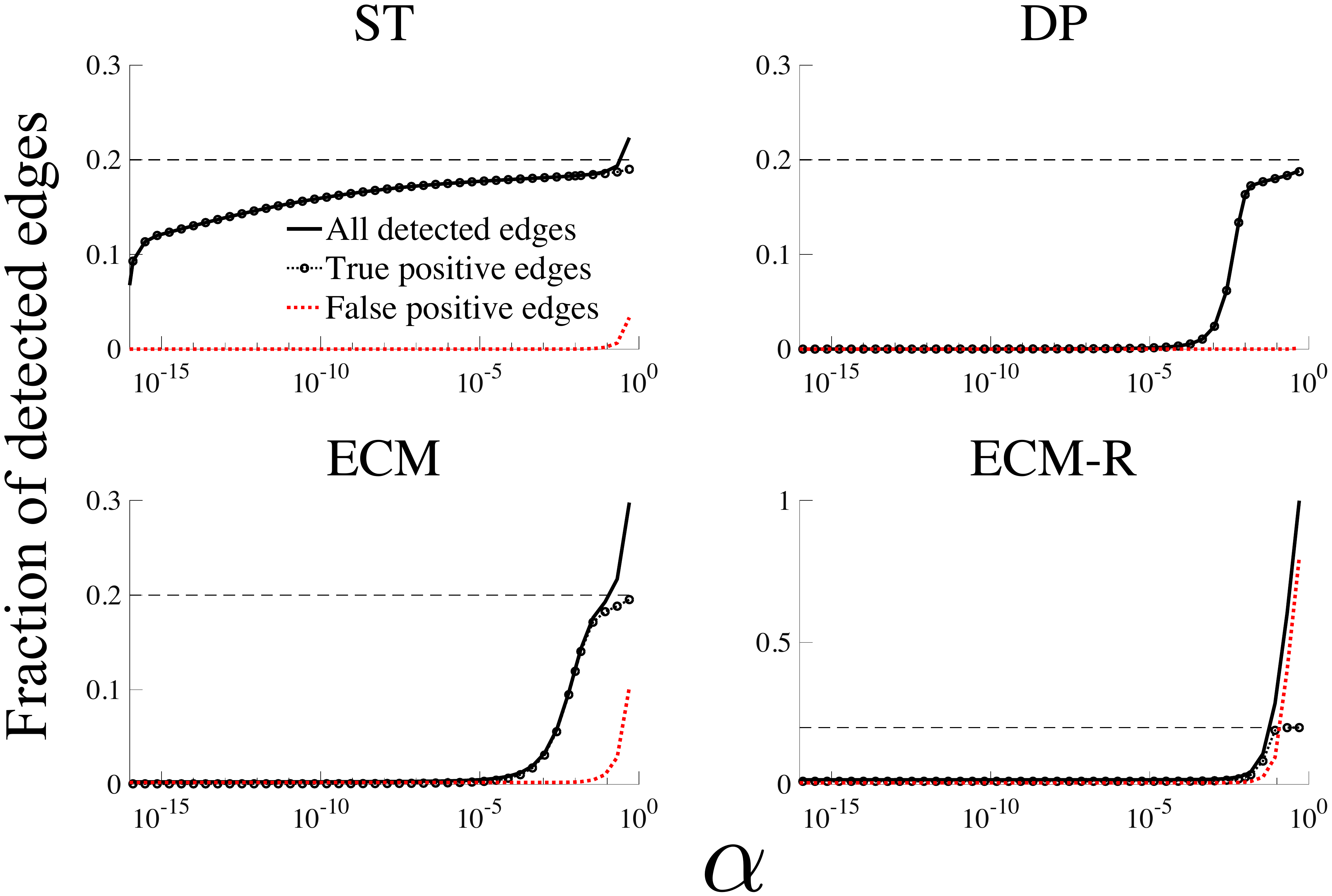}
    \end{center}
    \caption{\new{Fraction of detected edges vs. significance level for each filtering algorithm implemented on synthetic networks. The ground-truth fraction of strong edges is 0.2 (denoted by a black dashed line). The black solid lines denote the mean fraction of detected edges using the method given 
    at the top of each panel. The 
    dotted lines with circles illustrate the mean fraction of detected edges excluding false positive ones; i.e., the fraction contains only those edges that are genuinely strong in the synthetic data. Dotted red lines denote the mean fraction of detected significant ties 
    that are not strong ties in the synthetic data, i.e.,
    false positives.
    See section~\ref{sec:synthetic_process} for details about the generation of synthetic networks.       
    } }
 \label{fig:synthetic_excl}
\end{figure}

\new{

 To examine the detectability of strong ties in a controlled setting (section.~\ref{sec:synthetic}), we generate synthetic temporal networks in which the fraction of such ties is set a priori. 
The network-generating procedure is given by:
\begin{enumerate}
\item We consider $N$ nodes, each with an intrinsic
activity drawn from a Beta distribution, 
$a_i^{\prime}\sim {\rm Beta}(1,10)$.

\item We generate a temporal network in the time-window
$[0,T'-1]$: at each time-step, each pair of nodes $(i,j)$
is connected with probability $a_i^{\prime}a_j^{\prime}, \forall i\neq j$.
This yields a 
sequence of $T^\prime$ undirected and unweighted networks, $\widehat{A}(0),\widehat{A}(1),\ldots,\widehat{A}(T^{\prime}-1)$.

\item We will consider as ``data set" the last $T$ time-steps, i.e., 
the time-window $[T'-T, T'-1]$. Among the pairs with at least
one interaction in this time-window, we randomly select $20\%$
as having ``strong" ties.

\item We construct $T'$ new networks  
${A}(0), \cdots {A}(T'-1)$ from $\widehat{A}(0), \cdots \widehat{A}(T'-1)$ 
by adding interactions among the strong ties as
follows (for the other ties, we set $A_{ij}(t)= \widehat A_{ij}(t)$ for
$t=1,\cdots,T'-1$): for each strong tie $(i,j)$, we initialize
$A_{ij}(0)= \widehat A_{ij}(0)$ and we repeat for $t=1,\cdots,T'-1$

\begin{itemize}
    \item if $\widehat A_{ij}(t)=1$, we keep $A_{ij}(t)= 1$;
        \item if $\widehat A_{ij}(t)=0$ and $A_{ij}(t-1)=1$, i.e., 
        if $i$ and $j$ are ending an interaction, we set
        $A_{ij}(t)= 0$ with probability $h_{ij}(t)$ and
    $A_{ij}(t)= 1$ with probability $1-h_{ij}(t)$, where
$h_{ij}(t) = \frac{1}{1+b \cdot D_{ij}(t-1)}$
    and $D_{ij}(t-1)$ denotes the number of consecutive periods up to $t-1$ in which $i$ and $j$ are in interaction. In other terms, 
    the longer $i$ and $j$ have been interacting, the more probable it is
    that they continue to interact \cite{Zhao2011PRE}.
\end{itemize}
 The non-negative parameter $b$ tunes the strength of the strong ties; $b=0$ corresponds to a situation in which there are no strong ties.

\item We use the last $T$ time-steps as our synthetic data set: 
we create a sequence of $\tau=\lfloor T/\Delta\rfloor$ snapshots by aggregating over time-windows of $\Delta$ consecutive time-steps. 
In each snapshot, the aggregate edges are binarized for the ST filter and  
weighted by the number of interactions for the ECM-R filter. For the DP and ECM filters, a weighted network is created by aggregating over the $T$ snapshots. 
\end{enumerate}

We set $N=300$, $\Delta=10$, $T=300$ and $b=5$. A sequence of $T^{\prime} = 3000$ networks is generated in each run and the initial 2700 periods are discarded (i.e., $T = T^{\prime}-2700=300$). The average and the distributions shown in Figs.~\ref{fig:synthetic_net} and \ref{fig:synthetic_excl} are computed over 100 simulations.
}

\clearpage
\newpage

\section{Definitions of ROC curve and AUC} \label{sec:SI_ROC}

\begin{figure}[thb]
\begin{center}
           \includegraphics[width=.9\columnwidth]{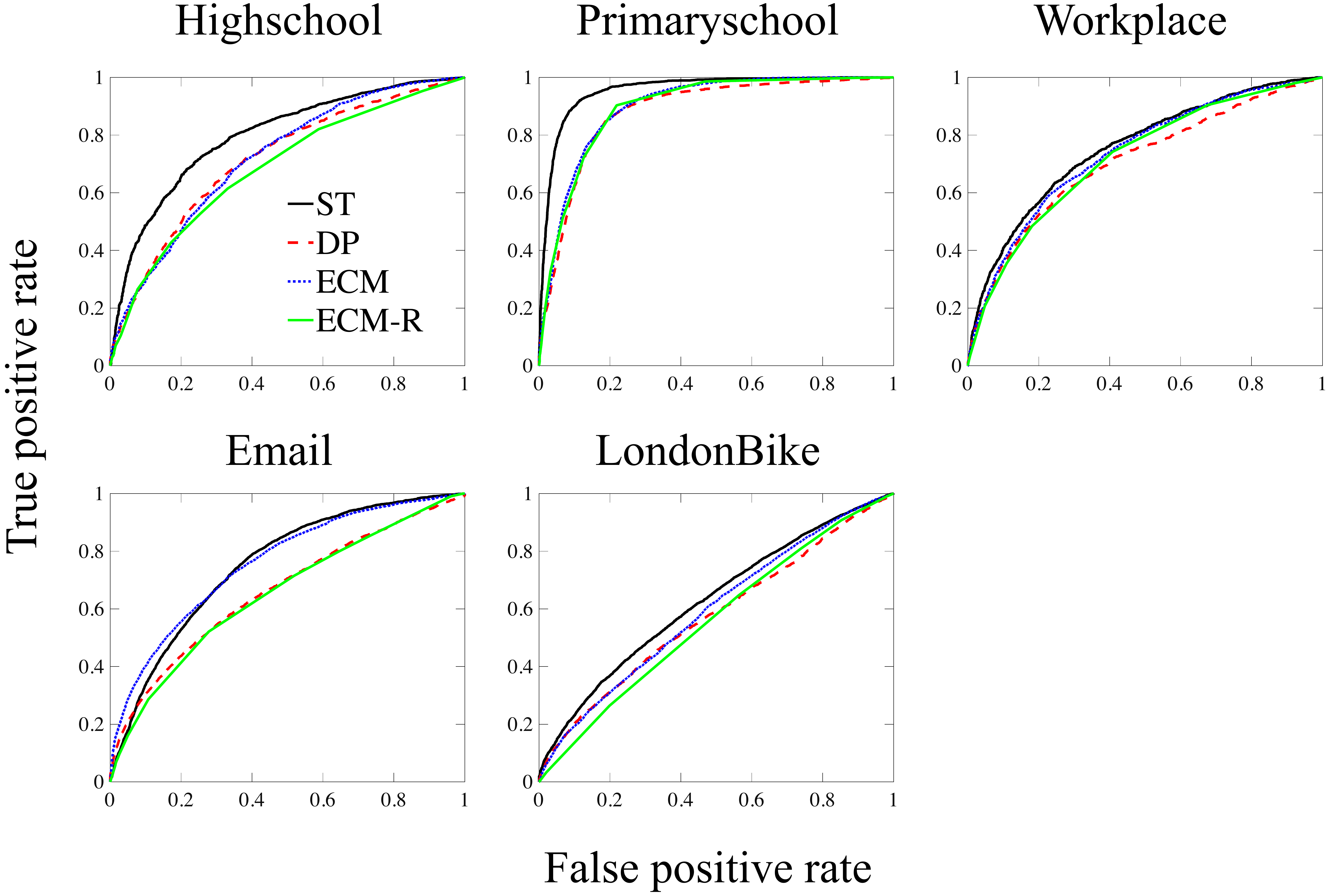}
    \end{center}
    \caption{ROC curve for the detection of intra-community edges for the data
    sets with a community structure, for the filters considered
    in the main text: Disparity filter (DP), Enhanced Configuration model (ECM),
    ECM-R and ST filter. 
    }
 \label{fig:ROC}
\end{figure}

 The receiver operating characteristic (ROC) curve is a plot of true positive rates against false positive rates for different cutpoints of a test statistic.  
 In our context, we want to know how well the significance of an edge can predict whether that edge is an intra-community edge.
For this purpose, we use the 
$p$-value of an edge in a given filtering test as a measure of edge significance. 
That is, different points on an ROC curve denote different cutoff levels of $p$-values for a given filtering method (Fig.~\ref{fig:ROC}).
For the ST filter presented in the main text, the $p$-value of an edge $i-j$ is
simply $1-G(m_{ij}^{c} |a_{i}^*,a_{j}^*)$.

For a given $p_0$, we consider therefore as True Positives (TP) the edges
that have a $p$-value lower than $p_0$ (i.e., are considered significant)
and are intra-community edges. False Positives (FP) are instead the
inter-community edges with a $p$-value lower than $p_0$.  
 Similarly, True Negatives (TM) are edges with $p$-value larger than $p_0$
 and inter-community, and False Negatives (FN) the intra-community edges
 with $p$-value larger than $p_0$. The false positive rate is given by $FP/(TN+FP)$
 and the true positive rate by $TP/(TP+FN)$.

 The area under the ROC curve (AUC) quantifies the goodness of the $p$-values for
 the task of predicting intra-community edges. 
 If the $p$-value of a filtering method
 perfectly distinguishes between intra- and inter-community edges
 (i.e., higher $p$-values indicate inter-community edges), then the value of AUC will be one.
 If the $p$-value is instead a poor indicator so that it is not different from a random prediction, then the AUC will be close to $0.5$ 
 (i.e., the ROC curve is then be a straight line
 joining the points $(0,0)$ and $(1,1)$). 
We show the ROC curves and a comparison of the AUCs for different data sets in
Fig.~\ref{fig:ROC} and Fig.~\ref{fig:fixed_edge_visual}{\sf c} in the main text.

\clearpage
\newpage

\begin{figure}[t]
\begin{center}
\includegraphics[width=.98\columnwidth]{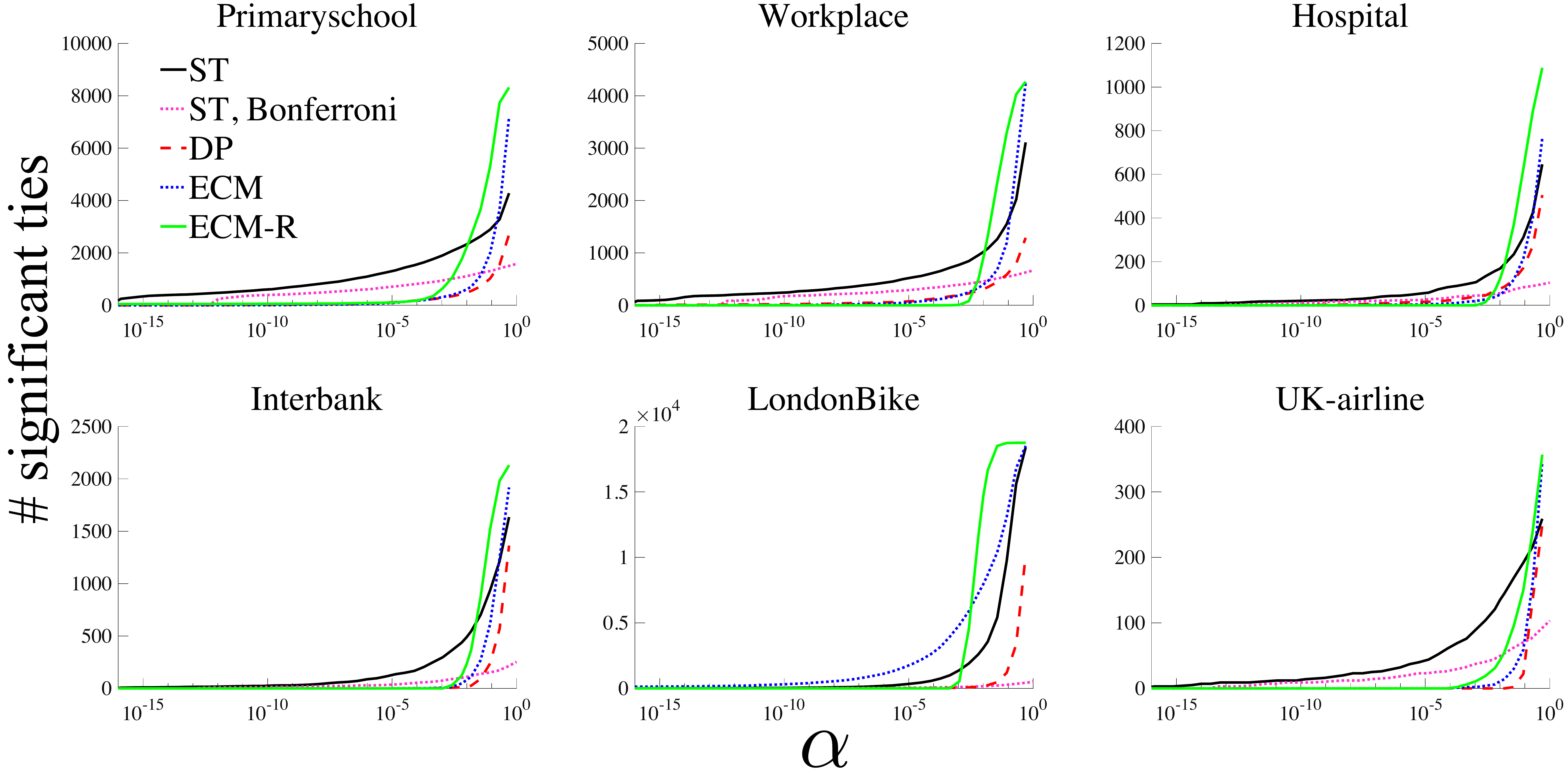}
    \end{center}
    \caption{Number of significant ties vs. $\alpha$, for the different filters considered. \new{See caption of Fig.~\ref{fig:number}}.}
 \label{fig:num_sigties_SI}
\end{figure}

\begin{figure}[t]
\begin{center}
\includegraphics[width=.9\columnwidth]{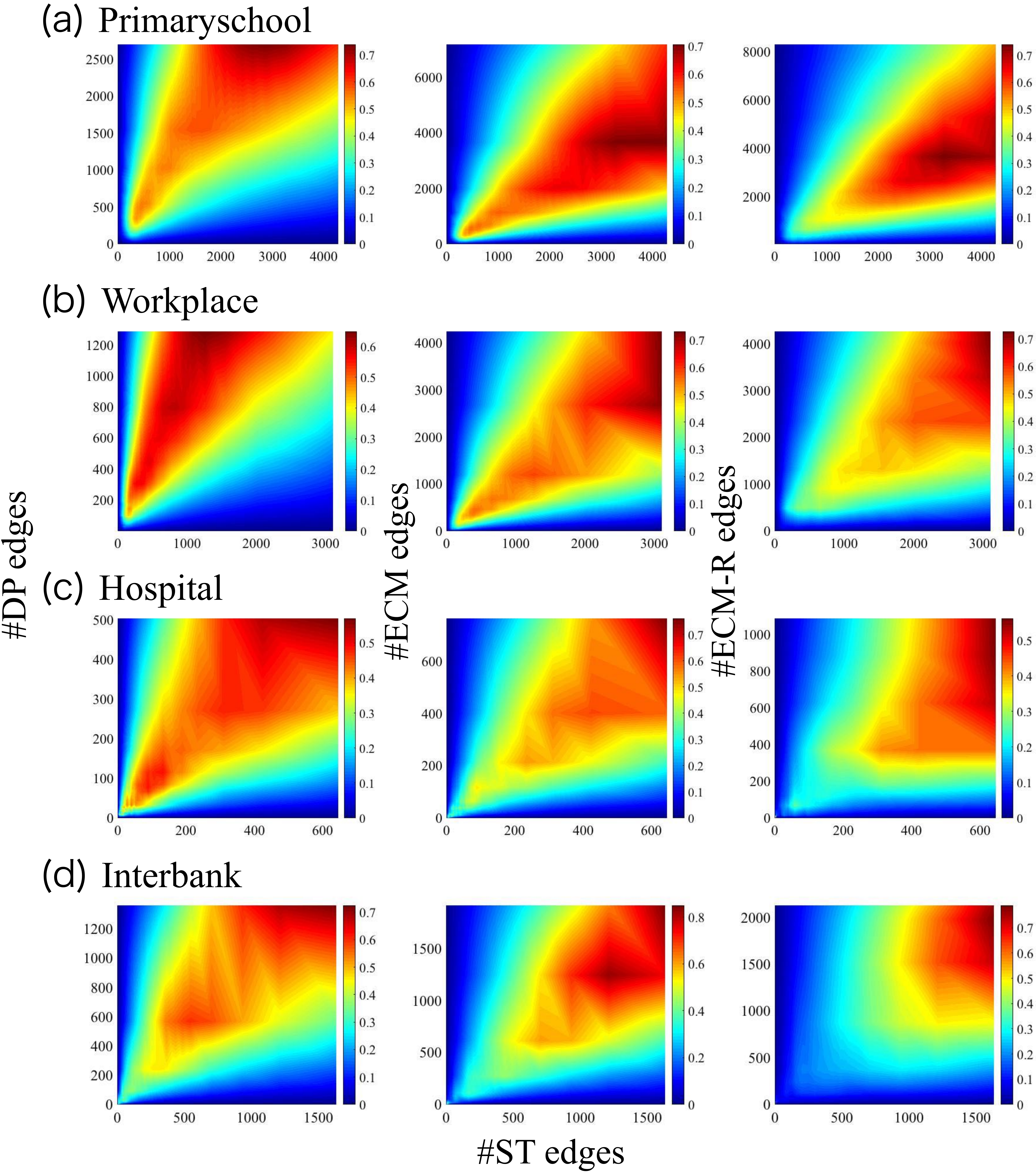}
    \end{center}
    \caption{Jaccard index of the similarity between the backbones obtained by various
    filtering methods, vs. the number of edges retained in each case. Jaccard index is defined by $J(I_{\rm ST}^\alpha,I_{\rm x}^{\alpha'}) = {|I_{\rm ST}^\alpha \cap I_{\rm x}^{\alpha'}|}/{|I_{\rm ST}^\alpha \cup I_{\rm x}^{\alpha'}|}$, where ${\rm x}$ denotes the filtering method \new{(DP, ECM, ECM-R)}.
 }
 \label{fig:Jaccard_SI}
\end{figure}
\begin{figure}[t]
\begin{center}
\includegraphics[width=.9\columnwidth]{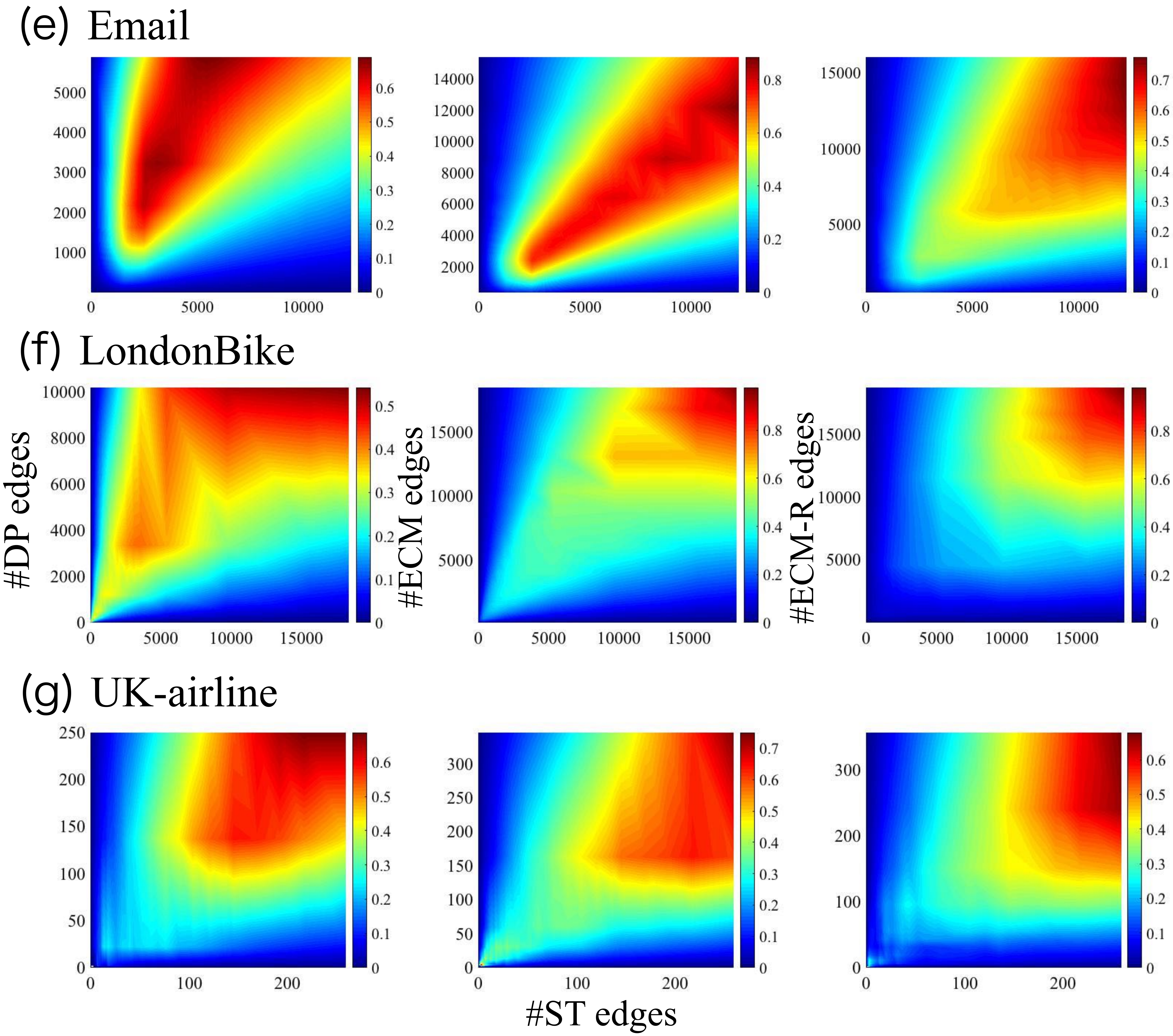}
    \end{center}
\end{figure}

\begin{figure}[t]
\begin{center}
\includegraphics[width=.9\columnwidth]{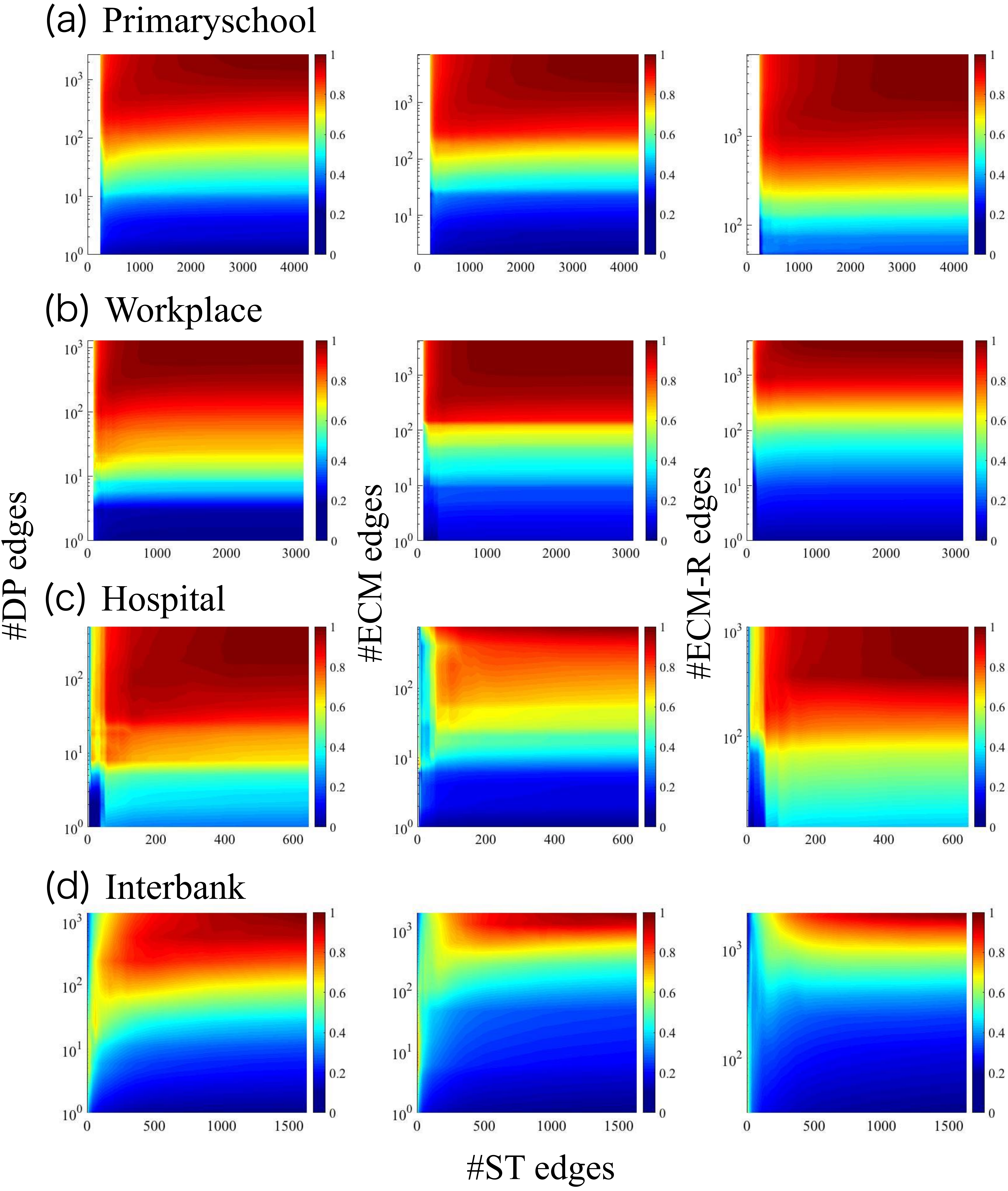}
    \end{center}
    \caption{Weighted measure of the similarity between the backbones obtained by various
    filtering methods, vs. the number of edges retained in each case. Here we use the
    cosine similarity between the weights of the edges retained by two methods, defined
    as $\sigma({\rm x},{\rm x'}) = \frac{\sum_{i<j}w_{ij}^{{\rm x}} w_{ij}^{{\rm x'}}}
  {\sqrt{\sum_{i<j}\left(w_{ij}^{{\rm x}}\right)^2}\sqrt{\sum_{i<j}\left(w_{ij}^{{\rm x'}}\right)^2}}$,
where ${\rm x}$ encodes both the filtering method \new{(ST, DP, ECM, ECM-R)} and the significance level $\alpha$ and the sums run
 on the pairs of nodes present in the backbones.}
 \label{fig:cosine_SI}
\end{figure}
\begin{figure}[t]
\begin{center}
\includegraphics[width=.9\columnwidth]{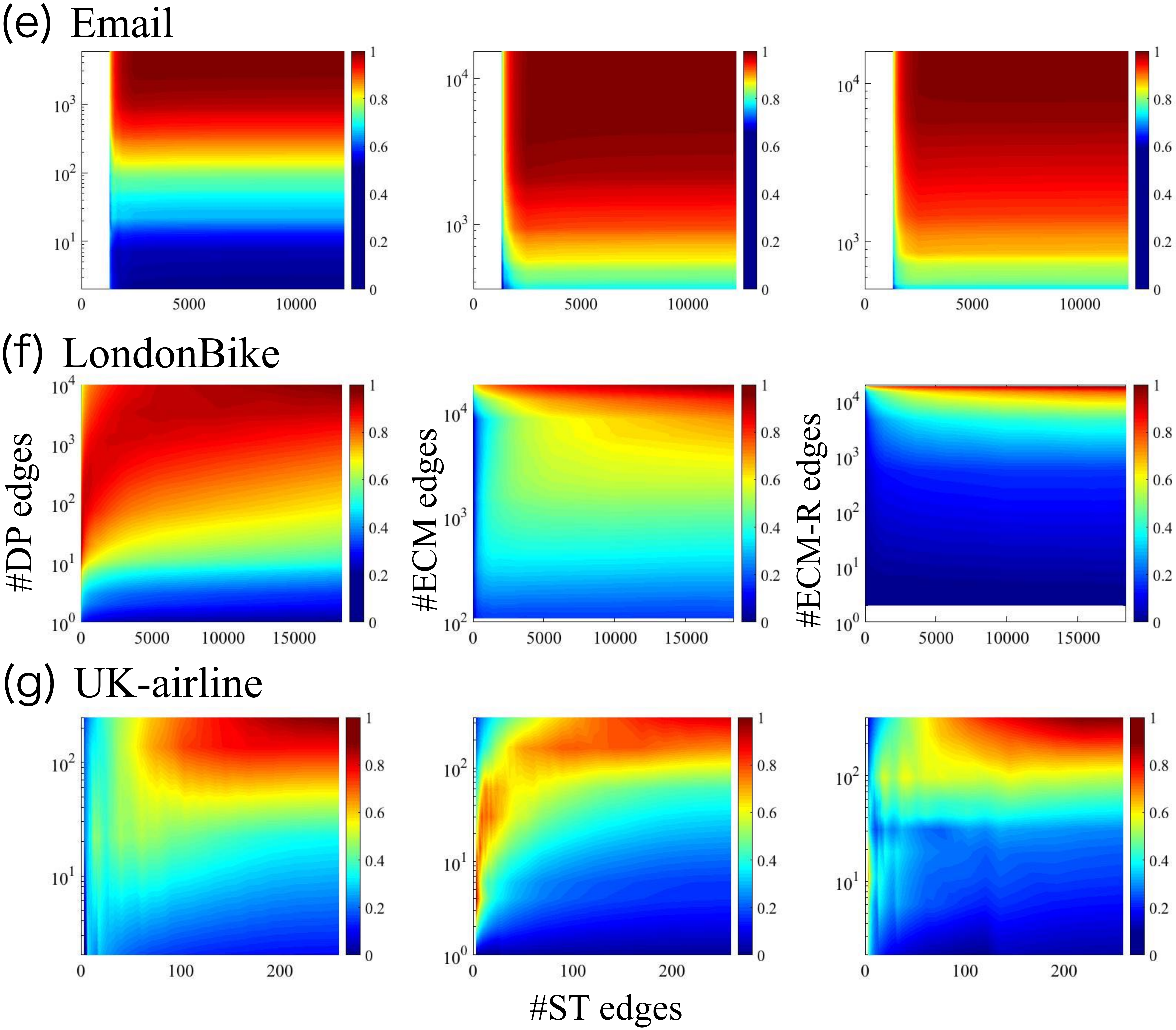}
    \end{center}
\end{figure}

\begin{figure}[t]
\begin{center}
\includegraphics[width=.9\columnwidth]{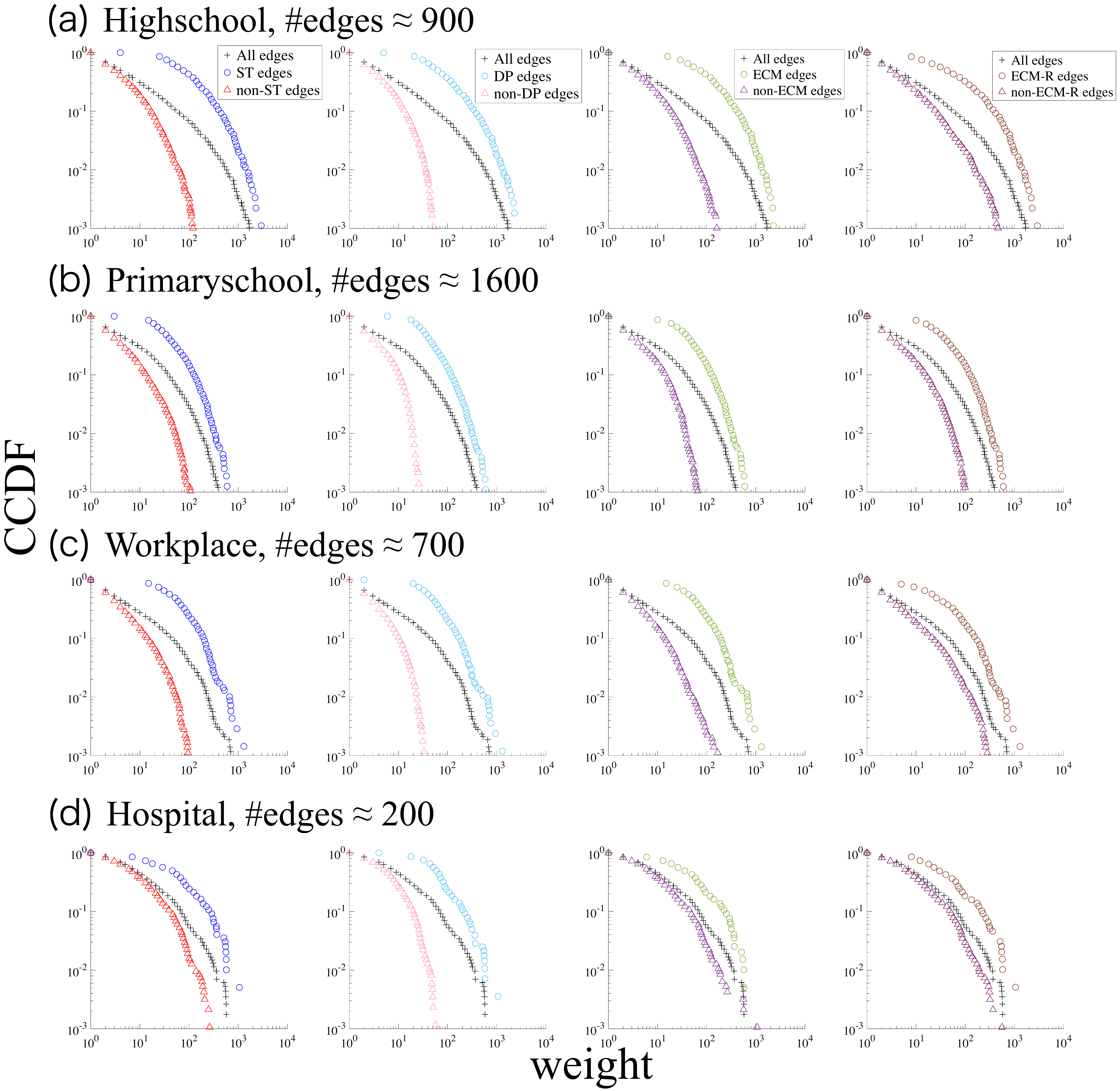}
    \end{center}
    \caption{\new{
    Weight distributions of significant and non-significant edges for a fixed number of edges, 
    for the ST filter (first
    column), the DP filter (second column), the ECM filter (third column) and the ECM-R filter (fourth column). In panels (a)--(g), the weights represent the total number of interactions. In panel (h), the total number of passengers is used as weight, since the number of interactions is quite small (namely, the number of flights is at most 14). DP is in general closer to a simple thresholding
    than the other filters (see discussion in the main text).
    }}
 \label{fig:weight_dist}
\end{figure}

\begin{figure}[t]
\begin{center}
\includegraphics[width=.9\columnwidth]{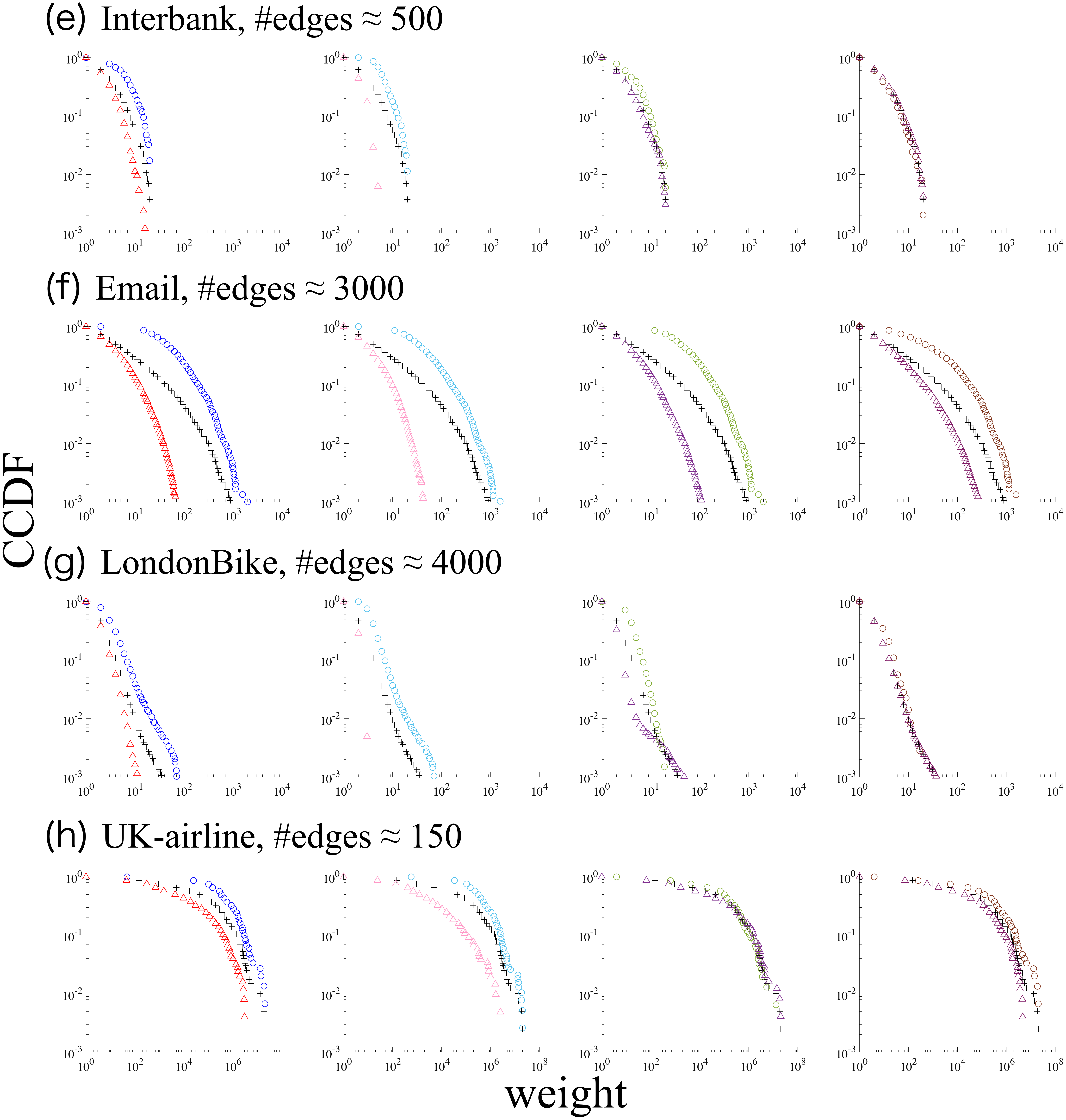}
    \end{center}
    \end{figure}

\begin{figure}[t]
\begin{center}
\includegraphics[width=.99\columnwidth]{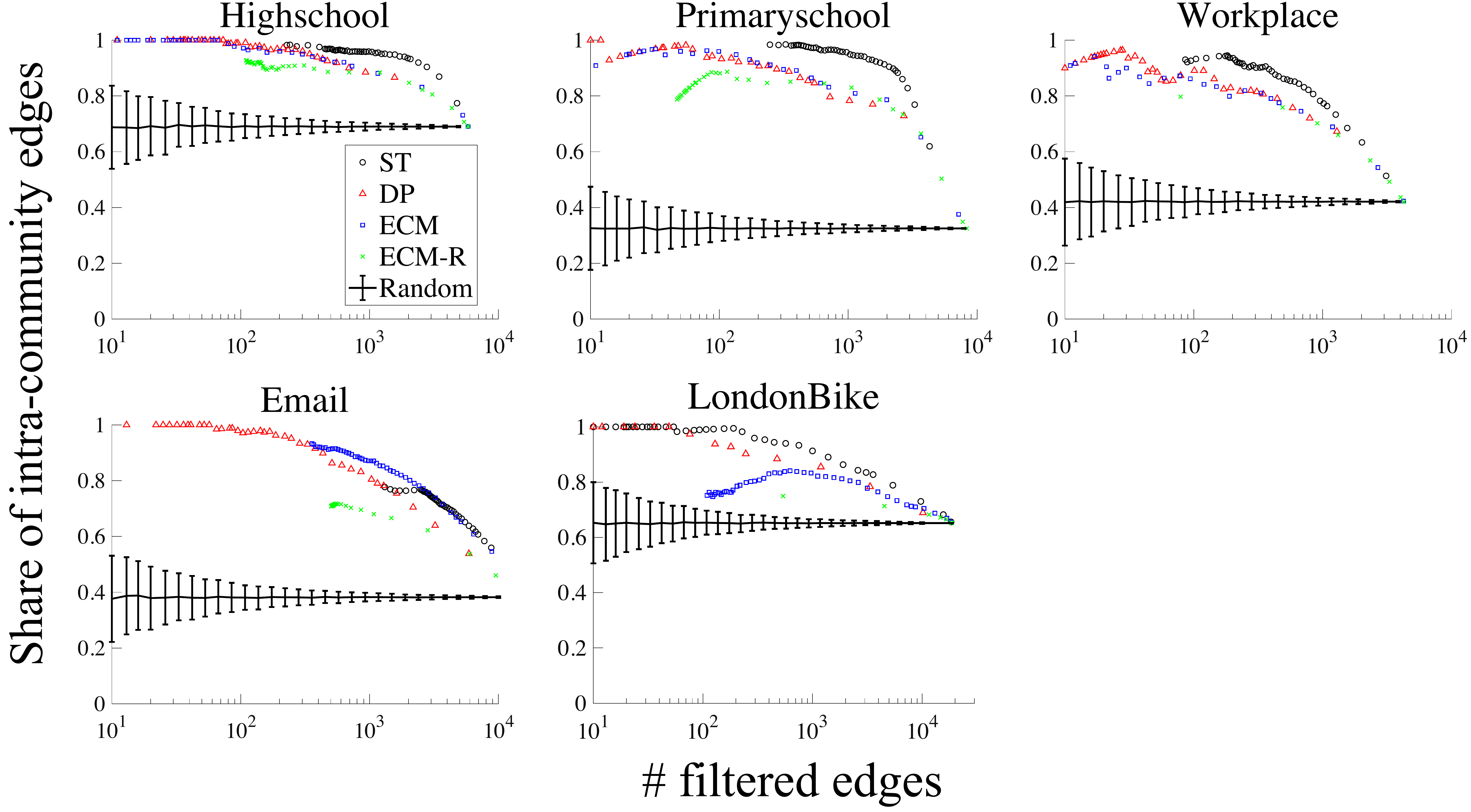}
    \end{center}
    \caption{\new{Share of intra-community edges among the significant edges, as a function of the number of such edges, for various filters. We consider the networks with $Q>0.3$. In addition to DP, ECM, ECM-R and ST we show here a random filter selecting edges at random. Error bar denotes the standard deviation calculated over 1,000 runs of random filtering.}
    }
 \label{fig:clust}
\end{figure}




\begin{figure}[t]
\begin{center}
\includegraphics[width=.98\columnwidth]{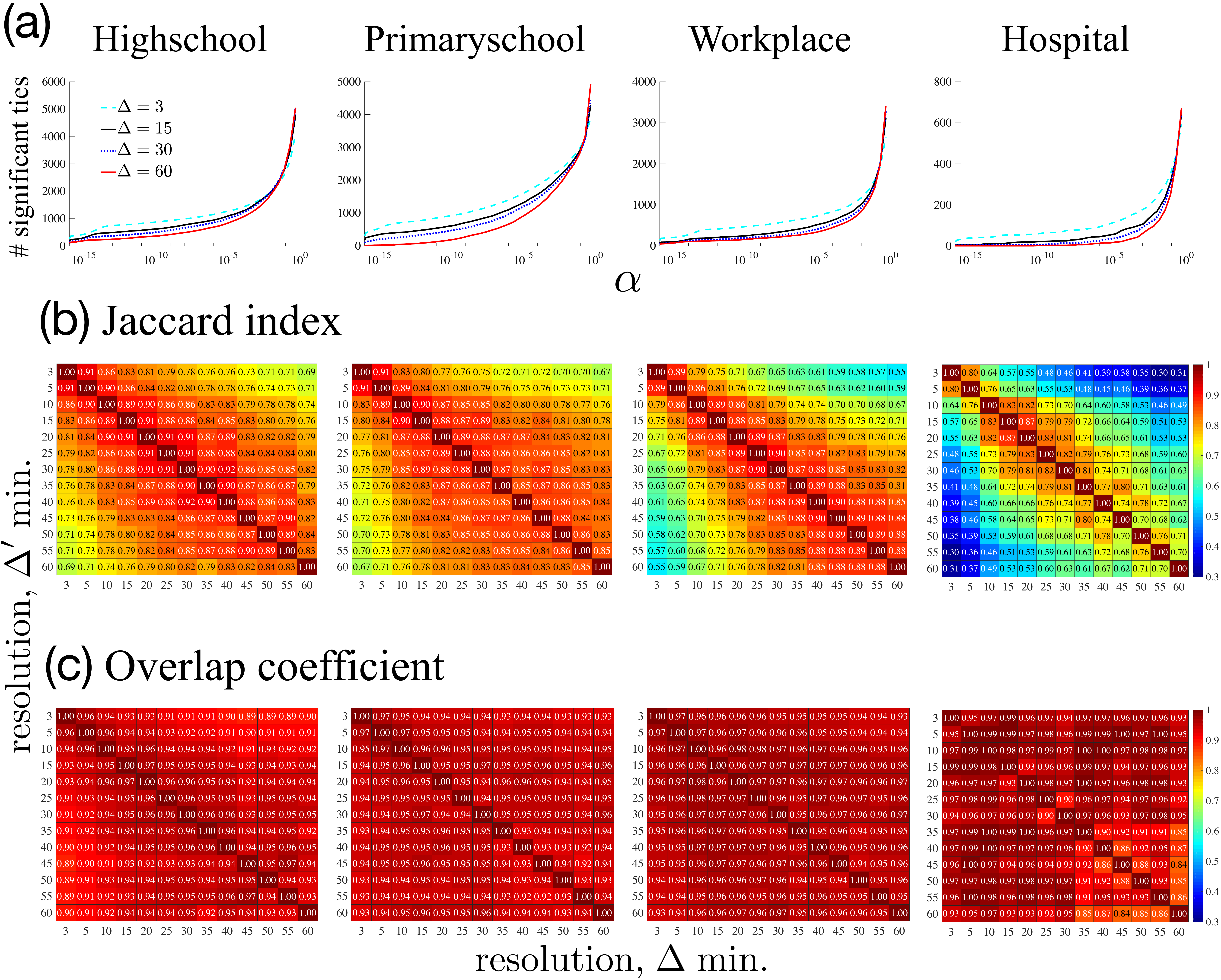}
    \end{center}
    \caption{\new{Significant ties at different temporal resolutions. (a) Number of significant ties at four different resolutions: $\Delta = \{3,15,30,60\}$. (b) and (c) Heatmap for the overlap of significant ties at different temporal resolutions. In panel (b), we consider the Jaccard index defined by $J(I_{\rm ST}^{\Delta},I_{\rm ST}^{\Delta^\prime}) \equiv | I_{\rm ST}^{\Delta} \cap I_{\rm ST}^{\Delta^\prime}|/| I_{\rm ST}^{\Delta} \cup I_{\rm ST}^{\Delta^\prime}|$, where $I_{\rm ST}^{\Delta}$ denotes the set of significant pairs detected at temporal resolution $\Delta$ minutes. In panel (c) we show instead the Overlap coefficient, or Szymkiewicz–Simpson coefficient, defined by $S(I_{\rm ST}^{\Delta},I_{\rm ST}^{\Delta^\prime}) \equiv | I_{\rm ST}^{\Delta} \cap I_{\rm ST}^{\Delta^\prime}|/ \min(| I_{\rm ST}^{\Delta}| , |I_{\rm ST}^{\Delta^\prime}|)$. In (b) and (c), the significance level is set at $\alpha=10^{-3}$. The difference between the two heatmaps indicates that the main difference stemming from temporal resolutions is just the number of detected pairs; namely, the set of significant ties detected at higher resolutions includes the significant ties detected at lower resolution. }}
 \label{fig:resol_heatmap}
\end{figure}

\begin{figure}[thb]
\begin{center}
\includegraphics[width=.99\columnwidth]{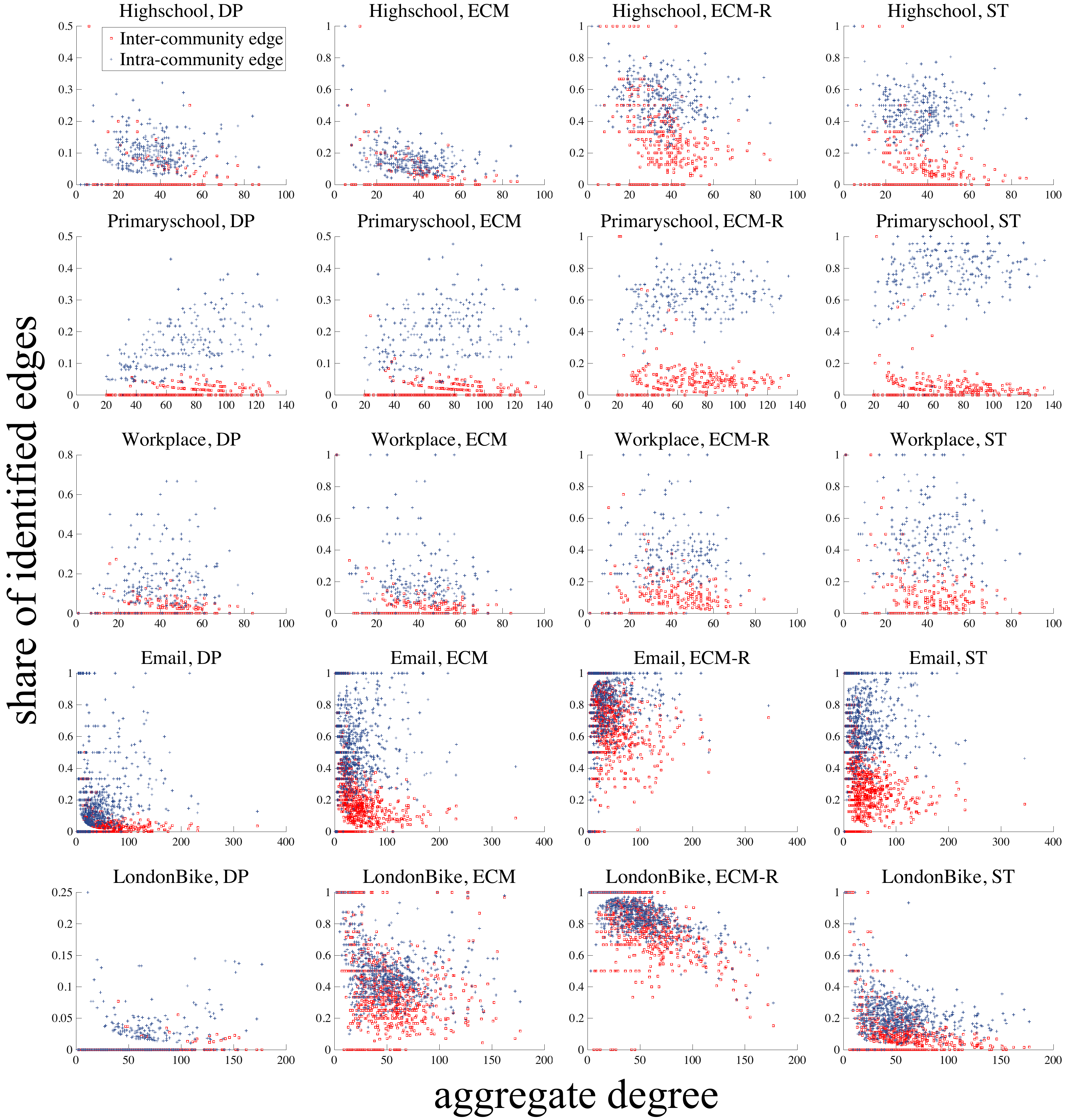}
    \end{center}
    \caption{\new{Share of filtered edges against aggregate degree in networks with community structure. The red squares (resp. the blue crosses) represent the share of significant inter-community (resp. intra-community) edges among all the inter-community (resp. intra-community) edges emanating from a node. Here
    $\alpha=0.01$
    The share of significant edges is generally larger among intra-community edges than among inter-community edges. 
    }}
 \label{fig:share_sig_inter}
\end{figure}

\begin{figure}[t]
\begin{center}
\includegraphics[width=.95\columnwidth]{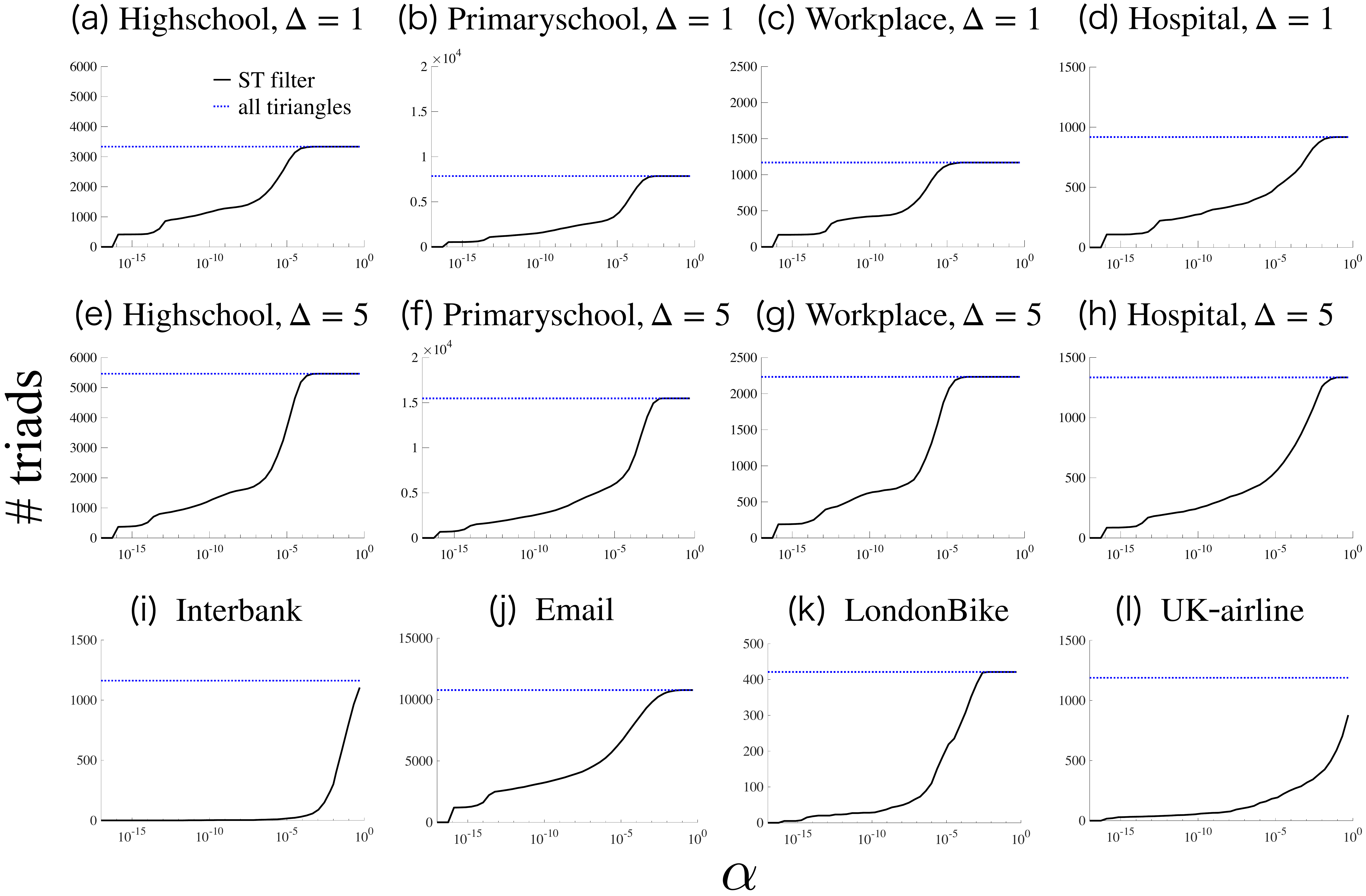}
    \end{center}
    \caption{Number of significant triads vs. the significance level $\alpha$, for
    different data sets and temporal resolutions.}
 \label{fig:num_triangle}
\end{figure}

\begin{figure}[t]
\begin{center}
\includegraphics[width=.95\columnwidth]{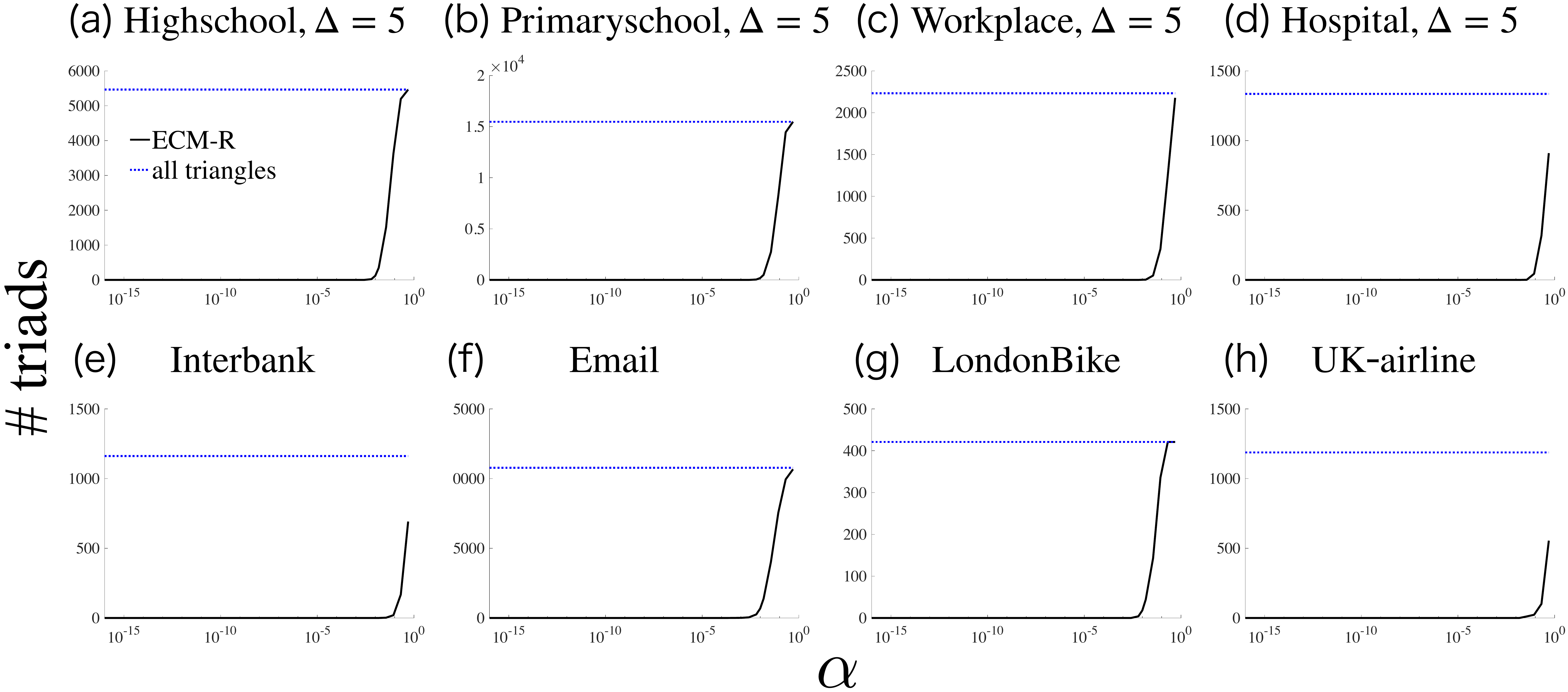}
    \end{center}
    \caption{\new{Number of triads formed by three ECM-R edges vs. the significance level $\alpha$. A triad is here regarded as ``significant" if the triad is formed by three ECM-R significant edges in at least one snapshot.}}
 \label{fig:num_triangle_ECMR}
\end{figure}

\begin{figure}[t]
\begin{center}
\includegraphics[width=.95\columnwidth]{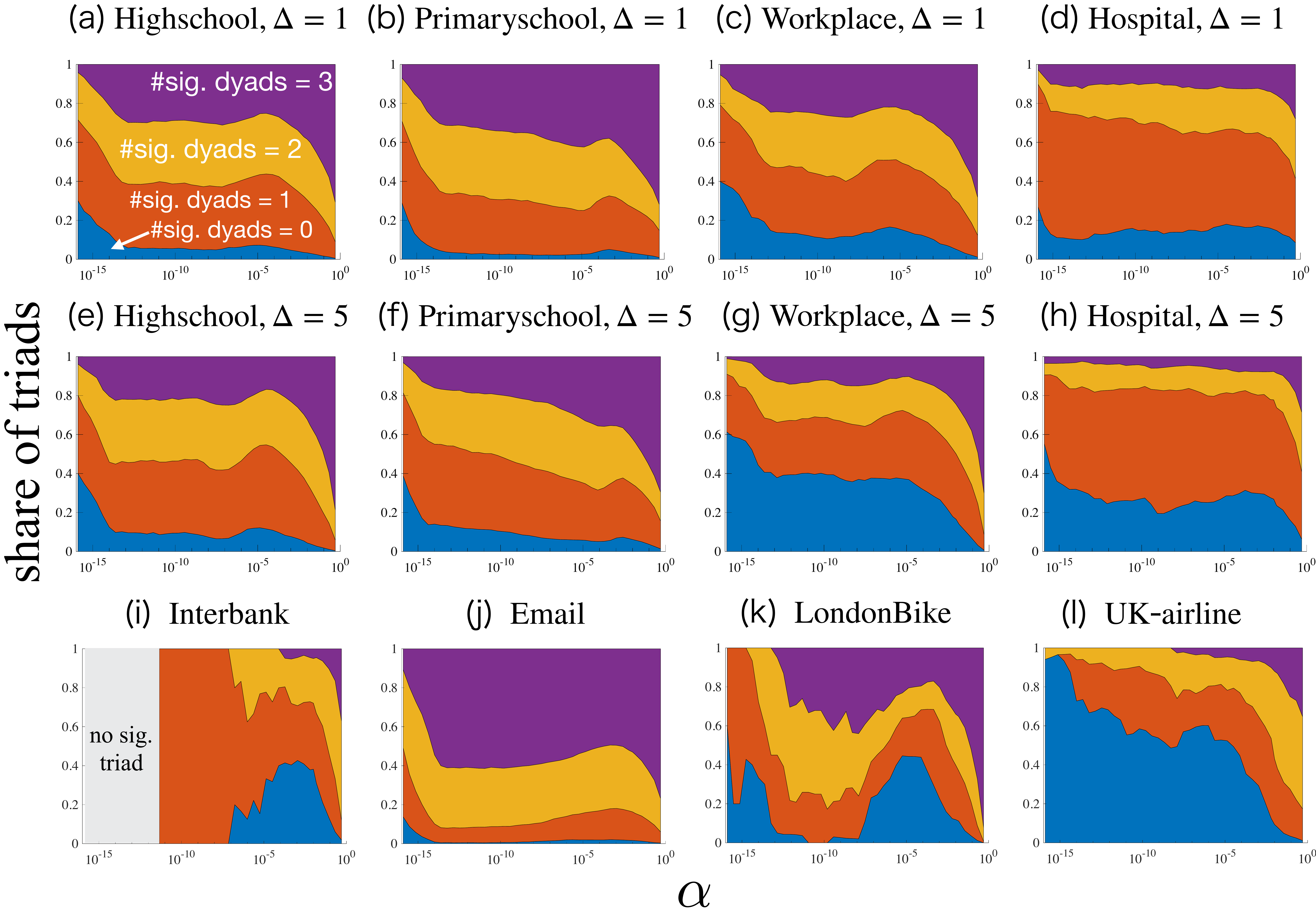}
    \end{center}
    \caption{Share of triangles with a given number of dyadic significant ties, vs. the
    filtering level $\alpha$, for various data sets and temporal resolutions.}
 \label{fig:share_triangle}
\end{figure}

\begin{figure}[t]
\begin{center}
\includegraphics[width=.95\columnwidth]{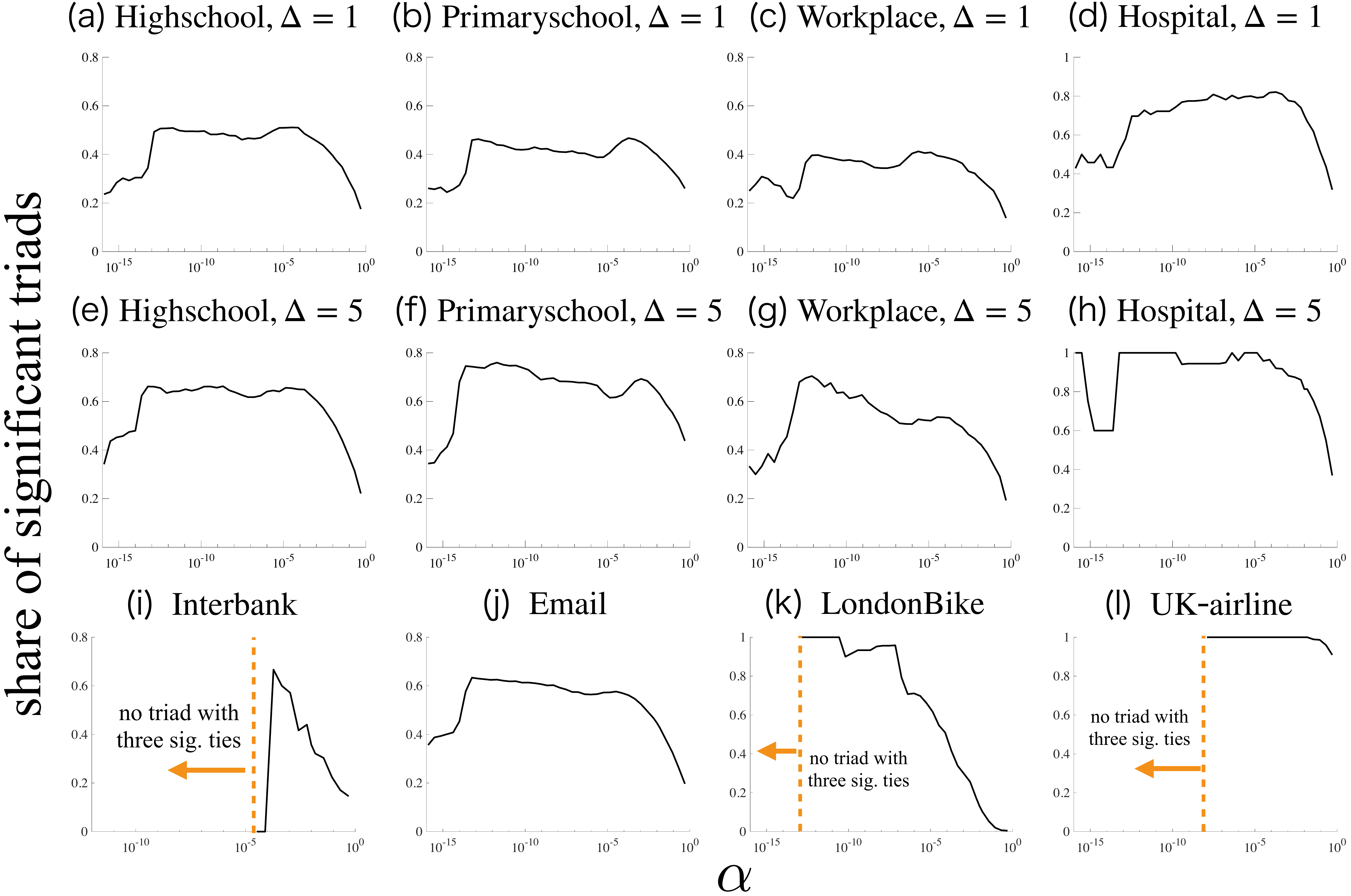}
    \end{center}
    \caption{Share of significant triads among the triangles composed by three significant ties vs. the
    filtering level $\alpha$, for various data sets and temporal resolutions.}
    \label{fig:frac_sigtri}
\end{figure}


\end{document}